\documentclass[12pt]{article}
\usepackage{amsmath,amsfonts}
\usepackage{float}
\usepackage{parskip}
\usepackage[letterpaper, margin=.75in]{geometry}
\usepackage{graphicx,caption,subcaption}
\usepackage{upgreek}
\usepackage{bm}
\usepackage{authblk}
\usepackage{wrapfig}
\usepackage[T1]{fontenc}
\usepackage[utf8]{inputenc}
\usepackage{siunitx}
\usepackage{textcomp}
\usepackage{xr-hyper} 
\usepackage{hyperref}
\usepackage{lineno}
\hypersetup{
    colorlinks=true,
    linkcolor=black,
    filecolor=black,      
    urlcolor=blue,
    citecolor=black,
}
 \makeatletter
 
\usepackage[backend=biber,style=nature,sorting=none]{biblatex}
\newcommand{\beginsupplement}{%
        \setcounter{table}{0}
        \renewcommand{\thetable}{S\arabic{table}}%
        \setcounter{figure}{0}
        \renewcommand{\thefigure}{S\arabic{figure}}%
     }

\makeatletter

\begin{document}

\title{Controlling Organization and Forces in Active Matter Through Optically-Defined Boundaries}

\author[1,2,*]{Tyler D. Ross}
\author[1,3]{Heun Jin Lee}
\author[1,2]{Zijie Qu}
\author[1,2]{Rachel A. Banks}
\author[1,2,3,4]{Rob Phillips}
\author[1,2,*]{Matt Thomson}
\affil[1]{California Institute of Technology, Pasadena, California, 91125, USA.}
\affil[2]{Division of Biology and Biological Engineering}
\affil[3]{Department of Applied Physics}
\affil[4]{Department of Physics}
\affil[*]{correspondence to: mthomson@caltech.edu, tross@caltech.edu}

\date{}
\maketitle

\begin{abstract}

Living systems are capable of locomotion, reconfiguration, and replication. To perform these tasks, cells spatiotemporally coordinate the interactions of force-generating, ``active'' molecules that create and manipulate non-equilibrium structures and force fields that span up to millimeter length scales \cite{Marchetti_review2013, dumont_emergent_2014,needleman_active_2017}. Experimental active matter systems of biological or synthetic molecules are capable of spontaneously organizing into structures \cite{nedelec_self-organization_1997,surrey_physical_2001} and generating global flows \cite{sanchez_spontaneous_2012,decamp_orientational_2015,wu_transition_2017,bricard_emergence_2013}. However, these experimental systems lack the spatiotemporal control found in cells, limiting their utility for studying non-equilibrium phenomena and bioinspired engineering. Here, we uncover non-equilibrium phenomena and principles by optically controlling structures and fluid flow in an engineered system of active biomolecules. Our engineered system consists of purified microtubules and light-activatable motor proteins that crosslink and organize microtubules into distinct structures upon illumination. We develop basic operations, defined as sets of light patterns, to create, move, and merge microtubule structures. By composing these basic operations, we are able to create microtubule networks that span several hundred microns in length and contract at speeds up to an order of magnitude faster than the speed of an individual motor. We manipulate these contractile networks to generate and sculpt persistent fluid flows. The principles of boundary-mediated control we uncover may be used to study emergent cellular structures and forces and to develop programmable active matter devices.

\end{abstract}

\begin{figure}[!htbp]
    \centering
    \includegraphics[width=.75\textwidth]{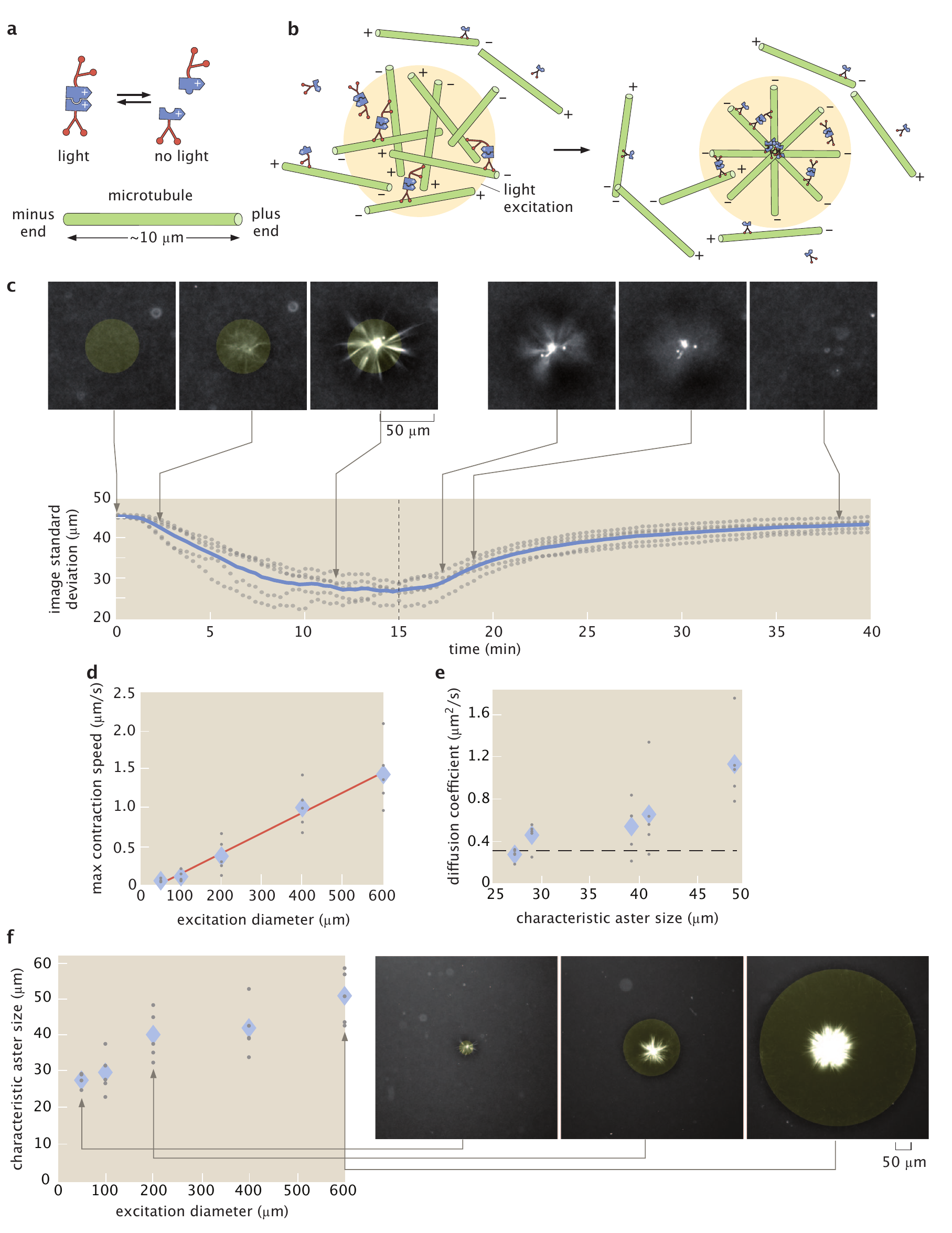}
    \caption{\label{fig:fig1} 
Light-switchable active matter system enables optical control over aster formation, decay and size. \textbf{a}, Schematic of light-dimerizable motors. \textbf{b}, Schematic of light-controlled reorganization of microtubules into an aster.  \textbf{c}, Images of labeled microtubules during aster assembly and decay and corresponding image spatial standard deviation versus time. The blue line is the mean of 5 experiments and the gray dots represent individual experiments. The dashed line is when the activation light is removed, transitioning from creation to decay. \textbf{d}, Max contraction speed versus excitation diameter. The red line is a linear fit. \textbf{e}, Diffusion coefficients versus characteristic aster size. The characteristic size is the image spatial standard deviation at the 15 minute time point shown in (\textbf{c}). The dashed line represents the diffusion coefficient of a 7 $\upmu$m microtubule (Supplementary Information~\ref{dsec:drag}).  \textbf{f}, Aster characteristic size versus excitation diameter with representative images. In (\textbf{d}, \textbf{e}, \textbf{f}) the diamonds represent the mean of 5 experiments and the gray dots represent individual experiments.  In (\textbf{c}, \textbf{f}), the yellow shaded disks represent the light pattern.}
\end{figure}

Our scheme is based on a well-studied active system composed of stabilized microtubule filaments and kinesin motor proteins \cite{nedelec_self-organization_1997,surrey_physical_2001,nedelec_dynamic_2001,lee_macroscopic_2001,sanchez_spontaneous_2012,decamp_orientational_2015,keber_topology_2014,wu_transition_2017}. In the original biochemical system, kinesin motors are linked together by practically irreversible biotin-streptavidin bonds. As linked motors pull on microtubules, a variety of phases and structures spontaneously emerge, such as asters, vortices, and networks. However, spatial and temporal control of these structures is limited \cite{surrey_physical_2001, aoyama_self-organized_2013}. 

We engineered the system so that light activates reversible linking between motors (Fig.~\ref{fig:fig1}a) by fusing Kinesin I motors to optically-dimerizable iLID proteins \cite{guntas_engineering_2015}. Light patterns are projected into the sample throughout its depth and determine when and where motors link (see Supplementary Information for details). Outside of the light excitation volume, microtubules remain disordered, while inside the light volume, microtubules bundle and organize. The reversibility of the motor linkages allows structures to remodel as we change the light pattern. For a cylinder pattern of light excitation, microtubules contract into a 3D aster (Fig.~\ref{fig:fig1}b) (Supplementary Information~\ref{dsec:3D}, \href{https://vimeo.com/327840168}{Video~1}, \href{https://vimeo.com/327840188}{Video~2}). We use the projection of a cylinder of light as an operation for creating asters. We note that vortices, spirals, and extensile behavior are not observed under our conditions (Supplementary Information~\ref{dsec:othersys}).

Our temporal control over aster formation allows us to study the dynamics of their creation and decay (Fig.~\ref{fig:fig1}c) (\href{https://vimeo.com/307204655}{Video~3}) through time lapse imaging (Supplementary Information~\ref{dsec:dataacquire}). We characterize these dynamics by measuring the spatial width of the distribution of fluorescently-labeled microtubules using image standard deviation (Supplementary Information~\ref{dsec:som}). During aster formation, the distribution of microtubules within a cylinder pattern contracts. After 10-15 min, the distribution reaches a steady state, indicating that the aster is fully formed. To quantify a characteristic aster size (Supplementary Information~\ref{dsec:charlength}), we measure the image standard deviation at 15 min (Supplementary Information~\ref{dsec:imageanalysis}). Once the excitation light is removed, asters begin to decay into free microtubules. The spatial distribution of microtubules widens over time, returning to the initial uniform distribution. Further, aster decay is reversible (Supplementary Information ~\ref{dsec:reversibility}).

To understand scaling behavior, we investigate how the dynamics of aster formation and decay depend on excitation volume. During formation, microtubule distributions contract. The contraction speed (Supplementary Information~\ref{dsec:form}) grows with the diameter of the excitation cylinder (Fig.~\ref{fig:fig1}d). Similar scaling of contraction speed has been observed for actomyosin systems \cite{schuppler_boundaries_2016} (Supplementary Information~\ref{dsec:NetworkComp}) and modeled for generic networks \cite{Belmonte_MSB_2017}. Alternatively, contraction can be measured by a characteristic contraction timescale  \cite{foster2015active} (see Supplementary Information~\ref{dsec:form}). During decay, microtubule distributions spread in a manner consistent with diffusion (Supplementary Information~\ref{dsec:decay}). The effective diffusion coefficient is independent of characteristic aster size (Fig.~\ref{fig:fig1}e) and is consistent with what is expected for free microtubules (Supplementary Information~\ref{dsec:drag}). Further, we manipulate aster size through the diameter of the excitation volume (Fig.~\ref{fig:fig1}f) and find a scaling dependence (Supplementary Information~\ref{dsec:asterscaling}) that shows similarities to the dependence of spindle size on confining volumes \cite{good_cytoplasmic_2013}. 

\begin{figure}[!htbp]
    \centering
    \includegraphics[width=.8\textwidth]{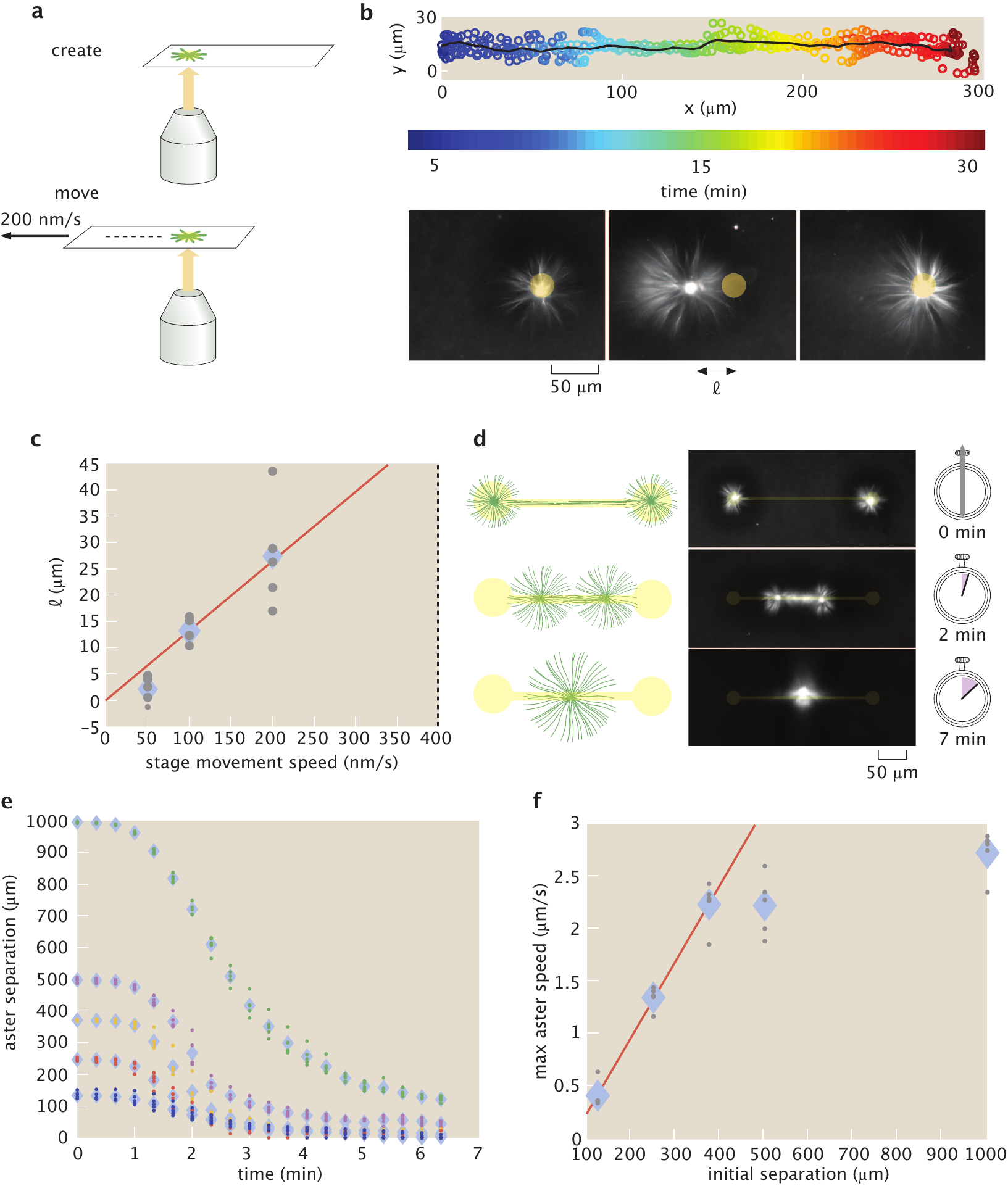}
    \caption{\label{fig:fig2} Moving and merging operations of asters with dynamic light patterns. \textbf{a}, Asters are moved relative to the slide by repositioning the microscope stage. \textbf{b}, Overlay of five individual trajectories of aster movement relative to slide moving at 200 nm/s. The line represents the mean trajectory. Time lapse images show the position of the aster relative to the light pattern. $\ell$ is the displacement of the aster from center of the light pattern. \textbf{c},  $\ell$  versus stage speed. The dotted line at 400 nm/s represents the escape velocity. The red line is a linear fit. \textbf{d}, Illustration of the aster merge operation by a connected excitation pattern and the corresponding time series of images. \textbf{e}, Distance between merging asters over time for different initial separations.  \textbf{f}, Maximum speeds of asters as measured from (\textbf{e}). The red line is a linear fit to the first three data points. In (\textbf{c}, \textbf{e}, \textbf{f}) the diamonds represent the mean of 5 experiments and the dots represent individual experiments.}
\end{figure}

Moving activation patterns are responsible for dynamically re-positioning structures and forces within a cell \cite{weiner_actin_2007}. We are able to similarly move asters by re-positioning light patterns relative to the sample slide by moving the slide stage (Fig.~\ref{fig:fig2}a). We are also able to move asters by directly moving the light pattern, however, moving the stage allows for a greater range of travel. As the stage moves, the asters track with the light pattern, traveling up to hundreds of microns relative to the slide (Fig.~\ref{fig:fig2}b) (\href{https://vimeo.com/307208989}{Video~4}) (Supplementary Information~\ref{dsec:astertrack}). The aster maintains a steady state distance $\ell$ between itself and the light pattern (Fig.~\ref{fig:fig2}c). We find that asters are always able to track the pattern for stage speeds up to 200 nm/s. At 400 nm/s asters are not able to stay with the pattern, setting an "escape velocity" that is comparable to the motor speeds measured in gliding assays (Supplementary Information~\ref{dsec:glidingspeed}). When the stage stops moving, the aster returns to the center of the light pattern, indicating that the aster is experiencing a restoring force. We can characterize aster movement as caused by an effective potential (Supplementary Information~\ref{dsec:movepotential}), and observe mesoscopic phenomena that may inform the underlying mechanisms of aster motion  (Supplementary Information~\ref{dsec:astermech}). 

Intriguingly, we find that asters formed near each other interact by spontaneously merging. To study this interaction, we construct an aster merger operation, where asters are connected with light (Fig.~\ref{fig:fig2}d) (\href{https://vimeo.com/307205769}{Video~5}). At the beginning of the merging process, a network of bundled microtubules forms, which connects the asters. The connecting network begins to contract and the asters move towards each other (Fig.~\ref{fig:fig2}e). The speed at which asters merge (Supplementary Information~\ref{dsec:form}) increases as a function of linking distance up to a speed of roughly 2.5 $\upmu$m/s (Fig.~\ref{fig:fig2}f). The scaling of aster merger speed as a function of distance is similar to the observed relationship of contraction speed as a function of the excitation cylinder size discussed above. We note that the maximum observed merger speed is about an order of magnitude higher than the speeds observed during gliding assays (Supplementary Information~\ref{dsec:glidingspeed}), which is analogous to how cell migration speeds can exceed single motor speeds \cite{Waterman_Migration2010}. Our ability to move and merge microtubule asters reveals that they are not steady state structures as previously observed \cite{surrey_physical_2001}, but are dynamic and constantly remodeling.

\begin{figure}[!hbtp]
    \centering
    \includegraphics[width=0.68\textwidth]{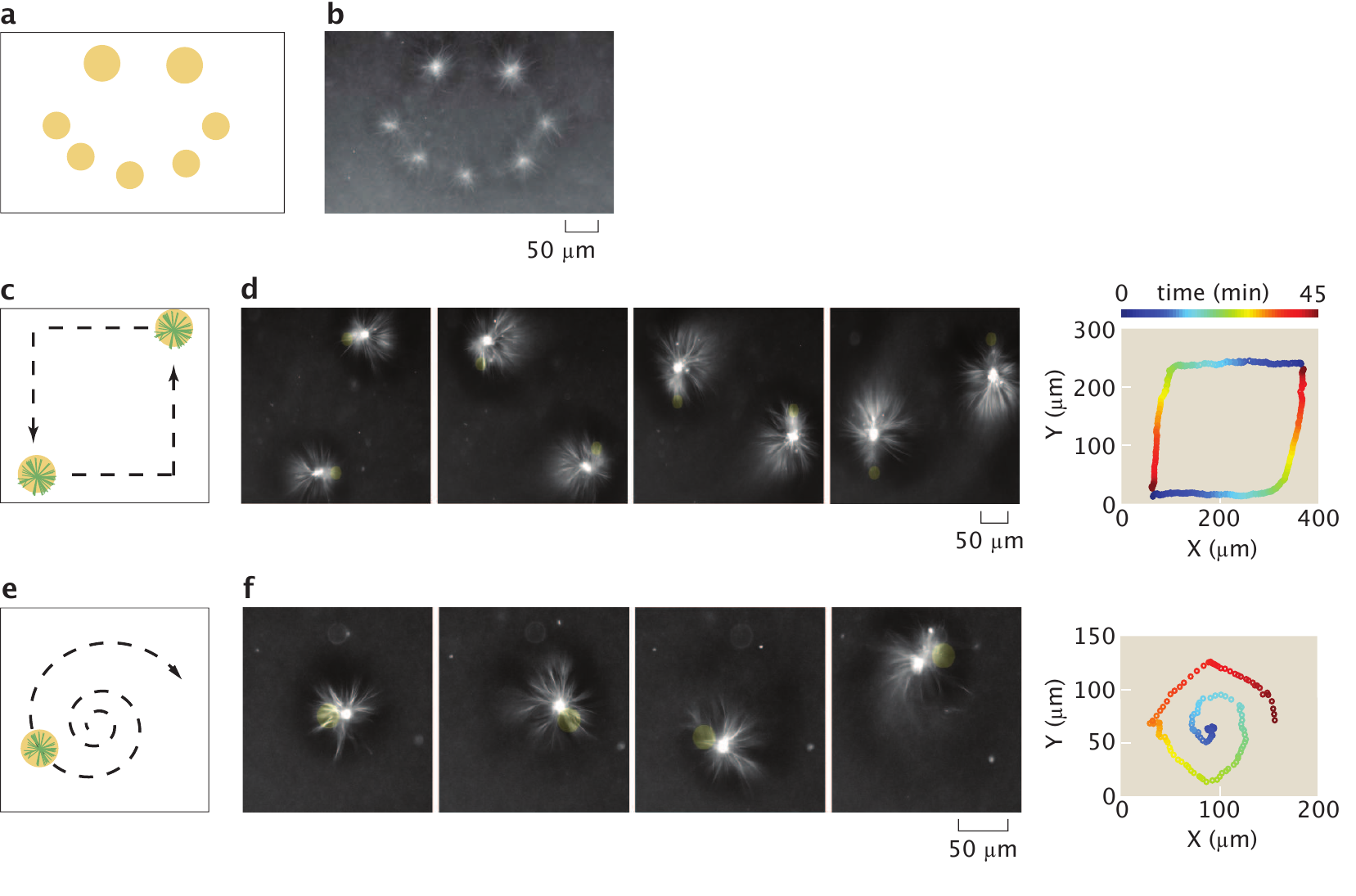}
    \caption{\label{fig:fig3} Operations for creating and moving asters are composed to make different desired patterns and trajectories. \textbf{a}, Sketch for using excitation cylinders to simultaneously pattern asters of different sizes.  \textbf{b}, Resultant pattern of asters corresponding to (\textbf{a}). \textbf{c}, Illustration of simultaneous control of two different aster trajectories, as indicated by the dashed arrows. \textbf{d}, Time lapse and the 2D trace of the aster trajectories corresponding to (\textbf{c}). The trajectory trace is color-coded to represent progression in time. \textbf{e}, Dynamically projected spiral to illustrate curvilinear motion. \textbf{f}, time lapse and the 2D trace of the aster trajectory. Time is color coded as in (\textbf{d}).}
\end{figure}

The capability to perform successive operations remains a fundamental step towards engineering with active matter. Our ability to form dynamic light-defined compartments of active molecules enables us to execute multiple aster operations. By composing aster creation operations, we are able to form asters of differing sizes and place them at prescribed positions in parallel (Fig.~\ref{fig:fig3}a, b) (\href{https://vimeo.com/307206914}{Video~6}). Once asters are created, they can be simultaneously moved by using multiple dynamic light patterns (Fig.~\ref{fig:fig3}c, d) (\href{https://vimeo.com/307204915}{Video~7}). Further, aster trajectories are not limited to rectilinear motion but can be moved along complex trajectories (Fig.~\ref{fig:fig3}e, f) (\href{https://vimeo.com/307204819}{Video~8}). During movement, there are inflows of microtubule bundles created in the light pattern, which feed into the aster. There are also outflows of microtubules, which appear as comet-tail streams following the asters (Fig.~\ref{fig:fig3}d, f). These mass flows illustrate some of the complex non-equilibrium dynamics that are introduced by moving boundaries of molecular activity. The new capability to simultaneously generate and manipulate asters provides a basis for ``programming'' complex systems of interacting non-equilibrium structures.

\begin{figure}[!htbp]
    \centering
    \includegraphics[width=0.68\textwidth]{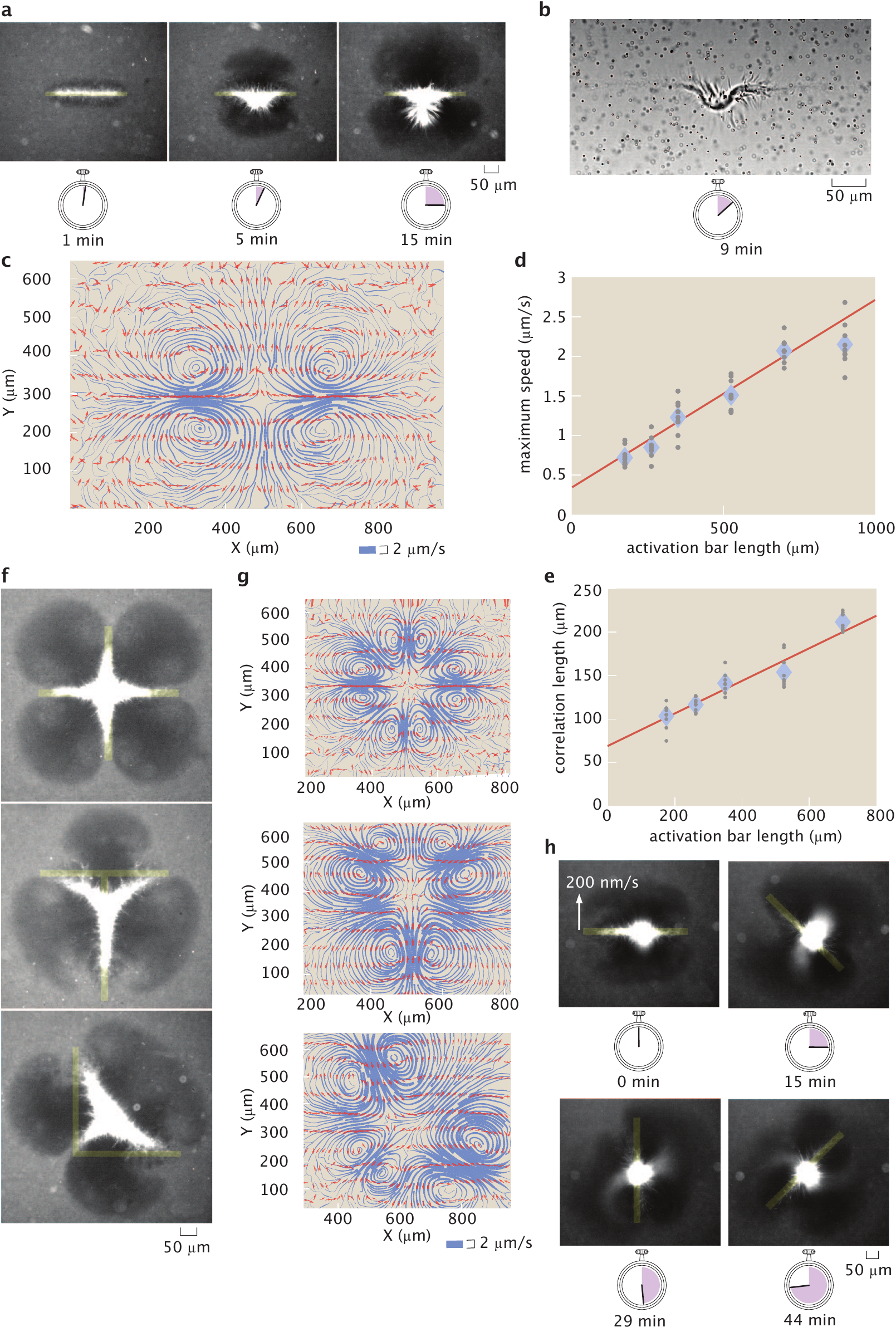}
    \caption{\label{fig:fig4} Advective fluid flow is created and controlled with patterned light. \textbf{a}, Microtubule organization created by an activation bar that is a 350 $\upmu$m x 20 $\upmu$m rectangular light pattern. Time series demonstrate continuous contraction of microtubules towards the pattern center along the major axis. \textbf{b}, Brightfield image of (\textbf{a}) shows a contracting microtubule network and tracer particles used to measure fluid flow. \textbf{c}, Streamline plots of background buffer flow from (\textbf{a}). The streamline thickness represents the flow speed. The arrows indicate the flow direction. \textbf{d}, Averaged maximum flow speed versus activation bar length. \textbf{e}, Averaged correlation length (size) of flow field versus activation bar length. \textbf{f}, Superposition of activation bars generate different patterns of contractile microtubules. \textbf{g}, Corresponding streamline plots. \textbf{h}, Time lapse of a light pattern rotating with an edge speed of 200 nm/s. In (\textbf{d}, \textbf{e}) the diamonds represents the mean of 9 experiments and the gray dots represent individual experiments. The red line is a linear fit to the data.}
\end{figure}

In our aster merging, moving, and trajectory experiments, we observe fluid flow of the buffer, as inferred by the advection of microtubules and small fluorescent aggregates. Similar cytoskeletal-driven flow is critical for the development and morphogenesis of various unicellular and multicellular organisms \cite{theurkauf1994premature, ganguly2012cytoplasmic, goldstein2008microfluidics, drescher2011fluid,drescher2010direct, he_apicalconstriction_2014, shinar_2011_modelcytoplasmically}.

Based on these observations, we seek to generate and tune flows in our engineered system with light, which may also provide insight into the mechanics of cellular fluid flow. Recent work has used light to thermally induce cytoplasmic flows \cite{Mittasch2018}. Here, we can generate fluid flows with light by activating contractile microtubule networks with the rectangular bar pattern used during aster merging (Fig.~\ref{fig:fig4}a) (\href{https://vimeo.com/307205580}{Video~9}). Brightfield images reveal a structurally changing microtubule network  (Fig.~\ref{fig:fig4}b) (\href{https://vimeo.com/307205646}{Video~10}), which appears to drive the fluid flow. We observe there are minimum size and angle limits for these microtubule structures, as well as for asters (Supplementary Information~\ref{dsec:minsize}).

We measure the flow fields with tracer particles (Supplementary Information~\ref{dsec:flowtrack}). The pattern of the flow is 2D (Supplementary Information~\ref{dsec:2Dflow}) and stable throughout the experiment (Supplementary Information~\ref{dsec:flowpatterns}), consisting of inflows and outflows of microtubules, as illustrated by streamline plots (Fig.~\ref{fig:fig4}c)(Supplementary Information~\ref{dsec:streamline}). The competition of these flows ensures that microtubules do not continuously accumulate in the illuminated region and that the surrounding medium is not completely depleted of microtubules.

We manipulate the properties of the flow field through the geometry of the activation volume. The size (Supplementary Information~\ref{dsec:correlation}) and speed of the flow field depend linearly on the length of the activation bar (Fig.~\ref{fig:fig4}d, e). The scaling of the flow speed is similar to the relationships for both the formation rate versus activation diameter and the aster merging speed versus separation. The positioning and number of inflows, outflows, and vortices are determined by the extrema of the light pattern geometry (Fig.~\ref{fig:fig4}f, g) (\href{https://vimeo.com/307206819}{Video~11}, \href{https://vimeo.com/307206846}{Video~12}, \href{https://vimeo.com/307206832}{Video~13}). A model that uses a series of point forces following the observed microtubule networks is able to recreate similar inflows and outflows (Supplementary Information~\ref{dsec:stokeslets}), suggesting that forces from microtubule bundles drive the flow.

Furthermore, the shape of the flow field has a temporal dependence on the light pattern. We modulate the flow field to create an ``active stir bar'' by applying a rotating light pattern (Fig.~\ref{fig:fig4}h) (\href{https://vimeo.com/307205942}{Video~14}). While simplified active matter systems are able to spontaneously generate global flows \cite{sanchez_spontaneous_2012, wu_transition_2017}, \textit{in vivo} cytoskeletal-driven fluid flows can be controlled and highly structured \cite{theurkauf1994premature, ganguly2012cytoplasmic, he_apicalconstriction_2014}. Our results demonstrate the creation and dynamic manipulation of localized, structured fluid flow in an engineered active matter system for the first time.

In this work, we uncover active matter phenomena through the creation and manipulation of non-equilibrium structures and resultant fluid flows. Our ability to define boundaries of protein activity with light enables unprecedented control of an active matter system's organization (Supplemental Information~\ref{dsec:swimsys}). We find scaling rules of contractile networks, movement of non-equilibrium structures, and modulation of flow fields. This framework may be built upon to create active matter devices that control fluid flow. Future work will explore spatiotemporal limits of non-equilibrium structures, the interplay of mass flows and structural changes, and develop new theories of non-equilibrium mechanics and dynamics. Our approach of understanding through construction creates a path towards a generalizable theory of non-equilibrium systems, engineering with active matter, and understanding biological phenomena.

\pagebreak

\textbf{Data Availability}
The data that support the findings of this study are available from the Caltech Research Data Repository: https://data.caltech.edu/records/1160. All plasmids used in this study are available on Addgene. All other reagents and source code used for this study are available from the corresponding author upon reasonable request.  

\textbf{Acknowledgements}
The authors would like to thank Maya Anjur-Dietrich, John Brady, Jehoshua Bruck, Vahe Galstyan, Soichi Hirokawa, Christina Hueschen, Yuri Lazebnik, Wendell Lim, Wallace Marshall, Dyche Mullins, Dan Needleman, Paul Rothemund, and Erik Winfree for influential scientific discussions. We thank Lukasz Bugaj, Zvonimir Dogic, Adam Frost, Walter Huynh, Rustem Ismagilov, Linnea Metcalf, Henry Nguyen, and Ron Vale for advice and assistance during development of the experimental system. Koen van den Dries for assistance with 3D visualization of asters. Paul Sternberg for use of a microscopy system for initial light activation experiments. We are grateful to Nigel Orme for assistance with figures and illustrations. The authors would like to acknowledge support from the NIH through grants 1R35 GM118043-01 (RP) and NIH DP5 OD012194 (MT); the NSF through NSF 1330864 (MT); the John Templeton Foundation as part of the Boundaries of Life Initiative Grants 51250 $\&$ 60973 (RP); The Foundational Questions Institute and Fetzer Franklin Fund  through FQXi 1816 (RP, MT); and the UCSF Center for Systems and Synthetic Biology NIGMS P50 GM081879 (MT).

\textbf{Author Contributions}
T.D.R., H.L., R.P, and M.T. conceived the experiments and interpreted the results. T.D.R., H.L., R.A.B., Z.Q., and M.T. wrote the manuscript. T.D.R. designed and cloned iLID motor fusion constructs. T.D.R., H.L., and R.A.B. performed protein purification. T.D.R. and H.L. designed, performed, and analyzed active matter experiments. Z.Q. analyzed and modeled flow data and tracked trajectories of moving asters. R.A.B. performed and analyzed gliding assays. All authors discussed results and commented on the manuscript.

\textbf{Competing Interests}
The authors declare no competing interests.

\pagebreak

\beginsupplement

\section{Methods and Materials}
\subsection{Kinesin Chimera Construction and Purification}\label{msec:construct}
To introduce optical control, we implemented the light-induced hetero-dimer system of iLID and SspB-micro \cite{guntas_engineering_2015}. We constructed two chimeras of \textit{D. melanogaster} kinesin K401:  K401-iLID and K401-micro (Fig S1).  

\begin{figure}[H]
    \centering
    \includegraphics[width=.8\textwidth]{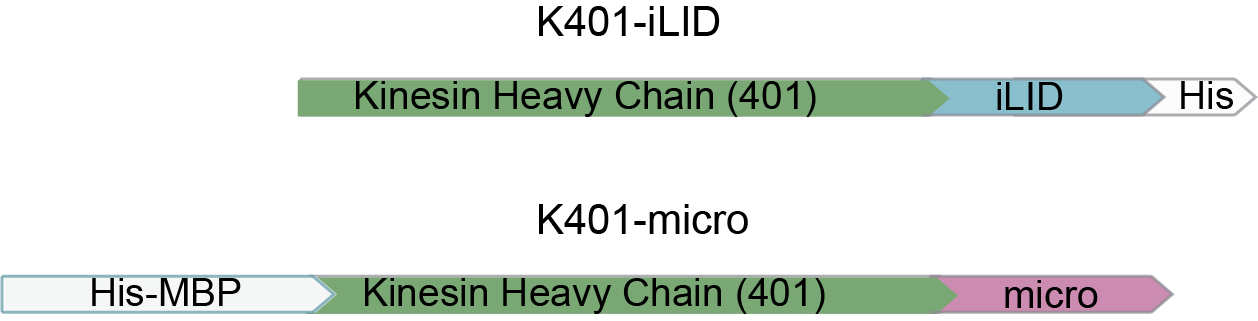}
    \caption{Kinesin motor coding regions}
    \label{fig:plasmid}
\end{figure}

To construct the K401-iLID plasmid (Addgene 122484), we PCR amplified the coding region of iLID from the plasmid pQE-80L iLID (gift from Brian Kuhlman, Addgene 60408) and used Gibson assembly to insert it after the C-terminus of K401 in the plasmid pBD-0016 (gift from Jeff Gelles, Addgene 15960).  To construct the K401-micro plasmid (Addgene 122485), we PCR amplified the coding region of K401 from the plasmid pBD-0016 and used Gibson assembly to insert it in between the His-MBP and micro coding regions of plasmid pQE-80L MBP-SspB Micro (gift from Brian Kuhlman, Addgene 60410).  As reported in \cite{guntas_engineering_2015}, the MBP domain is needed to ensure the micro domain remains fully functional during expression.  Subsequent to expression, the MBP domain can be cleaved off by utilizing a TEV protease site. 

For protein purification, we used the His tags that were provided by the base plasmids. For protein expression, we transformed the plasmids into BL21(DE3)pLysS cells.  The cells were induced at OD 0.6 with 1 mM IPTG and grown for 16 hours at 18\textdegree{}C.  The cells were pelleted and then resuspended in lysis buffer (50 mM sodium phosphate, 4 mM MgCl2, 250 mM NaCl, 25 mM imidazole, 0.05 mM MgATP, 5 mM BME, 1 mg/ml lysozyme and 1 tablet/50 mL of Complete Protease Inhibitor). After an hour, the lysate was passed through a 30 kPSI cell disruptor to lyse any remaining cells. The lysate was then clarified by an ultra-centrifuge spin at 30,000 g for 1 hour.  The clarified lysate was incubated with Ni-NTA agarose resin (Qiagen 30210) for 1 hour.  The lysate mixture was loaded into a chromatography column, washed three times with wash buffer (lysis buffer without lysozyme and protease inhibitor), and eluted with 500 mM imidazole. For the K401-micro elution, we added TEV protease at a 1:25 mass ratio to remove the MBP domain. Protein elutions were dialyzed overnight using a 30 kDa MWCO membrane to reduce trace imidazole and small protein fragments.  Protein was concentrated with a centrifugal filter (EMD Millipore UFC8030) to 8-10 mg/ml.  Protein concentrations were determined by absorption of 280 nm light with a UV spectrometer.                       

\subsection{Microtubule Polymerization and Length Distribution}\label{msec:polymerization}
We polymerized tubulin with the non-hydrolyzable GTP analog GMP-CPP, using a protocol based on the one found on the Mitchison lab homepage \cite{mitchison_protocol}. A polymerization mixture consisting of M2B buffer (80 mM K-PIPES pH 6.8, 1 mM EGTA, 2 mM MgCl2), 75 $\upmu$M unlabeled tubulin (PurSolutions 032005), 5 $\upmu$M tubulin-AlexaFluor647 (PurSolutions 064705), 1 mM DTT, and 0.6 mM GMP-CPP (Jenna Biosciences NU-405S) was spun at $\approx$ 300,000 g for 5 minutes at 2\textdegree{}C to pellet aggregates.  The supernatant was then incubated at 37\textdegree{}C for 1 hour to form GMP-CPP stabilized microtubules.    

\begin{figure}[H]
    \centering
    \includegraphics[width=.8\textwidth]{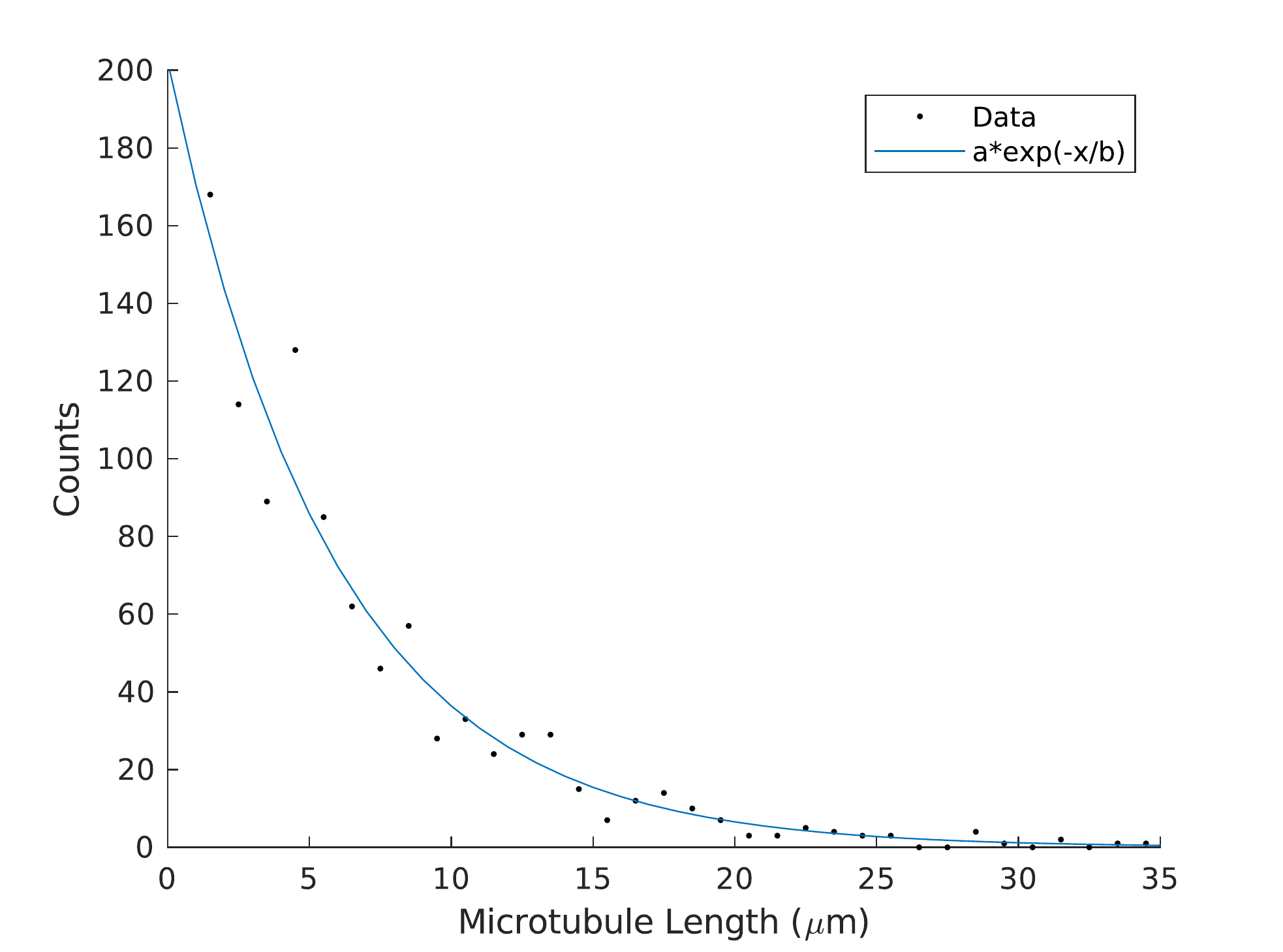}
    \caption{Length distribution of microtubules. The mean length given by the data histogram is $\ 7 \pm 0.2 \upmu$m, where the $\pm$ indicates the standard error of the mean. This mean length is similar to the $\approx 6 \upmu$m mean length given by a fit to an exponential distribution.}
    \label{fig:mtdist}
\end{figure}

To measure the length distribution of microtubules, we imaged fluorescently labeled microtubules immobilized onto the cover glass surface of a flow cell. The cover glass was treated with a 0.01\% solution of poly-L-lysine (Sigma P4707) to promote microtubule binding.  The lengths of microtubules were determined by image segmentation. To reduce the effect of the non-uniformity in the illumination, we apply a Bradley adaptive threshold with a sensitivity of 0.001 and binarize the image. Binary objects touching the image border and smaller than 10 pixels in size were removed. To connect together any masks that were ``broken'' by the thresholding, a morphological closing operation was performed with a 3 pixel $\times$ 3 pixel neighborhood. Masks of microtubules are then converted into single pixel lines by applying a morphological thinning followed by a removal of pixel spurs. The length of a microtubule is determined by counting the number of pixels that make up each line and multiplying by the interpixel distance. For the characteristic microtubule length, we report the mean of the measured lengths (Fig.~\ref{fig:mtdist}). For comparison, we also fit an exponential distribution to the observed histogram.  We note that a full distribution of microtubule lengths does not, in general, follow an exponential decay, however, the exponential has been shown to be appropriate for limited length spans \cite{Gardner_COCB_2013}. 

\subsection{Sample Chambers for Aster and Flow Experiments}\label{msec:flowcell} For the aster and flow experiments, microscope slides and cover glass are passivated against non-specific protein absorption with a hydrophilic acrylamide coating \cite{Lau_acrylamidecoat2009}. The glass is first cleaned in a multi-step alkaline etching procedure that removes organics and the surface layer of the glass. The slides and cover glass are immersed and sonicated for 30 minutes successively in 1\% Hellmanex III (Helma Analytics) solution, followed by ethanol, and finished in 0.1 M KOH solution.  After cleaning, the glass is immersed in a silanizing solution of 98.5\% ethanol, 1\% acetic acid, and 0.5\% 3-(Trimethoxysilyl)propylmethacrylate (Sigma 440159) for 10-15 min.  After rinsing, the slides are immersed overnight in a degassed 2 \% acrlylamide solution with 0.035\% TEMED and 3 mM ammonium persulfate. Just before use, the glass is rinsed in distilled water and nitrogen dried.  Parafilm M gaskets with pre-cut 3 mm wide channels are used to seal the cover glass and slide together, making a flow cell that is $\approx 70  \upmu$m in height.  After the addition of the reaction mixture, a flow cell lane is sealed with a fast setting silicone polymer (Picodent Twinsil Speed). 

\subsection{Reaction Mixture and Sample Preparation for Aster and Flow Experiments}\label{msec:reaction}
For the aster and flow experiments, K401-micro , K401-iLID , and microtubules were combined into a reaction mixture, leading to final concentrations of $\approx$ 0.1 $\upmu$M of each motor type and 1.5-2.5 $\upmu$M of tubulin. Concentrations refer to protein monomers for the K401-micro and K401-iLID constructs and the protein dimer for tubulin. To minimize unintended light activation, the sample was prepared under dark-room conditions, where the room light was filtered to block wavelengths below 580 nm (Kodak Wratten Filter No. 25).  The base reaction mixture provided a buffer, an energy source (MgATP), a crowding agent (glycerol), a surface passivating polymer (pluronic F-127), oxygen scavenging components to reduce photobleaching (glucose oxidase, glucose, catalase, Trolox, DTT), and ATP-recycling reagents to prolong motor activity (pyruvate kinase/lactic dehydrogenase, phosphoenolpyruvic acid).  The reaction mixture consisted of 59.2 mM K-PIPES pH 6.8, 4.7 mM MgCl2, 3.2 mM potassium chloride, 2.6 mM potassium phosphate, 0.74 mM EGTA, 1.4 mM MgATP (Sigma A9187),  10\% glycerol, 0.50 mg/mL pluronic F-127 (Sigma P2443), 0.22 mg/ml glucose oxidase (Sigma G2133), 3.2 mg/ml glucose, 0.038 mg/ml catalase (Sigma C40), 5.4 mM DTT, 2.0 mM Trolox (Sigma 238813),  0.026 units/$\upmu$l
pyruvate kinase/lactic dehydrogenase (Sigma P0294), and 26.6 mM phosphoenolpyruvic acid (Beantown Chemical 129745).

We note that the sample is sensitive to the ratio of motors and microtubules and the absolute motor concentration. When the motor concentration is below 0.1 $\upmu$M for K401-micro and K401-iLID, light patterns are able to create microtubule bundles or lattices of small asters, similar to the phases observed as functions of motor concentration described in \cite{surrey_physical_2001}. If this motor concentration is above $\approx$ 2 $\upmu$M, however, the number of binding events between inactivated K401-micro and K401-iLID proteins is sufficient to cause the spontaneous microtubule bundling and aster formation.

\subsection{Sample Preparation for Gliding Assay}\label{msec:gliding}
For the gliding assay experiments, microscope slides and cover glass are coated with antibodies to specifically bind motor proteins. First, alkaline cleaned cover glass and ethanol scrubbed slides were prepared and 5 $\upmu$L flow chambers were prepared with doubled sided tape. Motors were bound to the surface by successive incubations of the chamber with 400 $\upmu$g/mL penta-His antibody (Qiagen 34660) for 5 min, 10 mg/ml whole casein (Sigma C6554) for 5 min, and finally motor protein (1mg/mL in M2B) for 5 min. Unbound motors were washed out with M2B buffer, then AlexaFluor 647 labeled GMP-CPP stabilized microtubules in M2B with 5 mM MgATP and 1mM DTT were flowed in.  

\subsection{Preparation of Tracer Particles}\label{msec:tracerparticles}
To measure the fluid velocity, we used 1 $\upmu$m polystyrene beads (Polysciences 07310-15) as tracer particles. To passivate the hydrophobic surface of the beads, we incubated them overnight in M2B buffer with 50 mg/ml of pluronic F-127. Just before an experiment, the pluronic coated beads are washed by pelleting and resuspending in M2B buffer with 0.5 mg/ml pluronic to match the pluronic concentration of the reaction mixture.     

\subsection{Microscope Instrumentation}
We performed the experiments with an automated widefield epifluorescence microscope (Nikon TE2000). We custom modified the scope to provide two additional modes of imaging: epi-illuminated pattern projection and LED gated transmitted light.  We imaged light patterns from a programmable DLP chip (EKB TEchnologies DLP LightCrafter™ E4500 MKII™ Fiber Couple) onto the sample through a user-modified epi-illumination attachment (Nikon T-FL). The DLP chip was illuminated by a fiber coupled 470 nm LED (ThorLabs M470L3). The epi-illumination attachment had two light-path entry ports, one for the projected pattern light path and the other for a standard widefield epi-fluorescence light path.  The two light paths were overlapped with a dichroic mirror (Semrock BLP01-488R-25). The magnification of the epi-illuminating system was designed so that the imaging sensor of the camera (FliR BFLY-U3-23S6M-C) was fully illuminated when the entire DLP chip was on. Experiments were run with Micro-Manager \cite{micro-manager}, running custom scripts to controlled pattern projection and stage movement.  For the transmitted light path, we replaced the standard white-light brightfield source (Nikon T-DH) with an electronically time-gated 660 nm LED (ThorLabs M660L4-C5). This was done to minimize light-induced dimerization during bright field imaging.

\section{Data Acquisition, Analysis, and Supplemental Discussion}

\subsection{Aster Distribution in 3D}\label{dsec:3D}
\begin{figure}[H]
    \centering
    \includegraphics[width=0.8\textwidth]{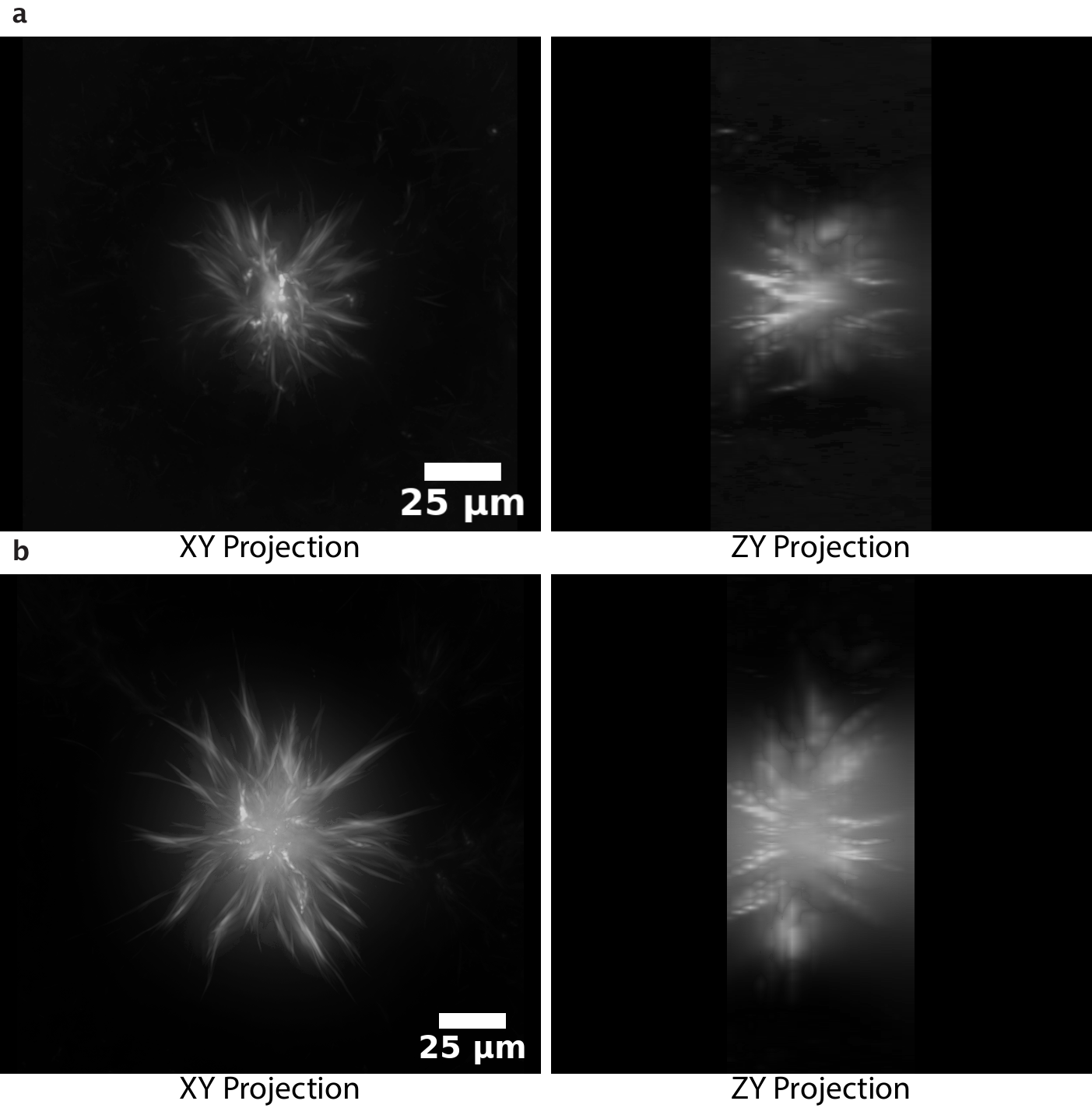}
    \caption{3D projections of asters from Z-stacks imaged with a 20x objective. \textbf{a}, Aster generated with a 100 $\upmu$m disk (\href{https://vimeo.com/327840168}{Video~1}). \textbf{b}, Aster generated with a 300 $\upmu$m disk (\href{https://vimeo.com/327840188}{Video~2}). The XY plane is along the plane of the sample slide. The ZY plane is orthogonal to the sample slide and the image is constructed by interpolating over 18 Z-slices spaced by 4 $\upmu$m. }
    \label{fig:3DAsters}
\end{figure}

\begin{figure}[H]
    \centering
    \includegraphics[width=0.8\textwidth]{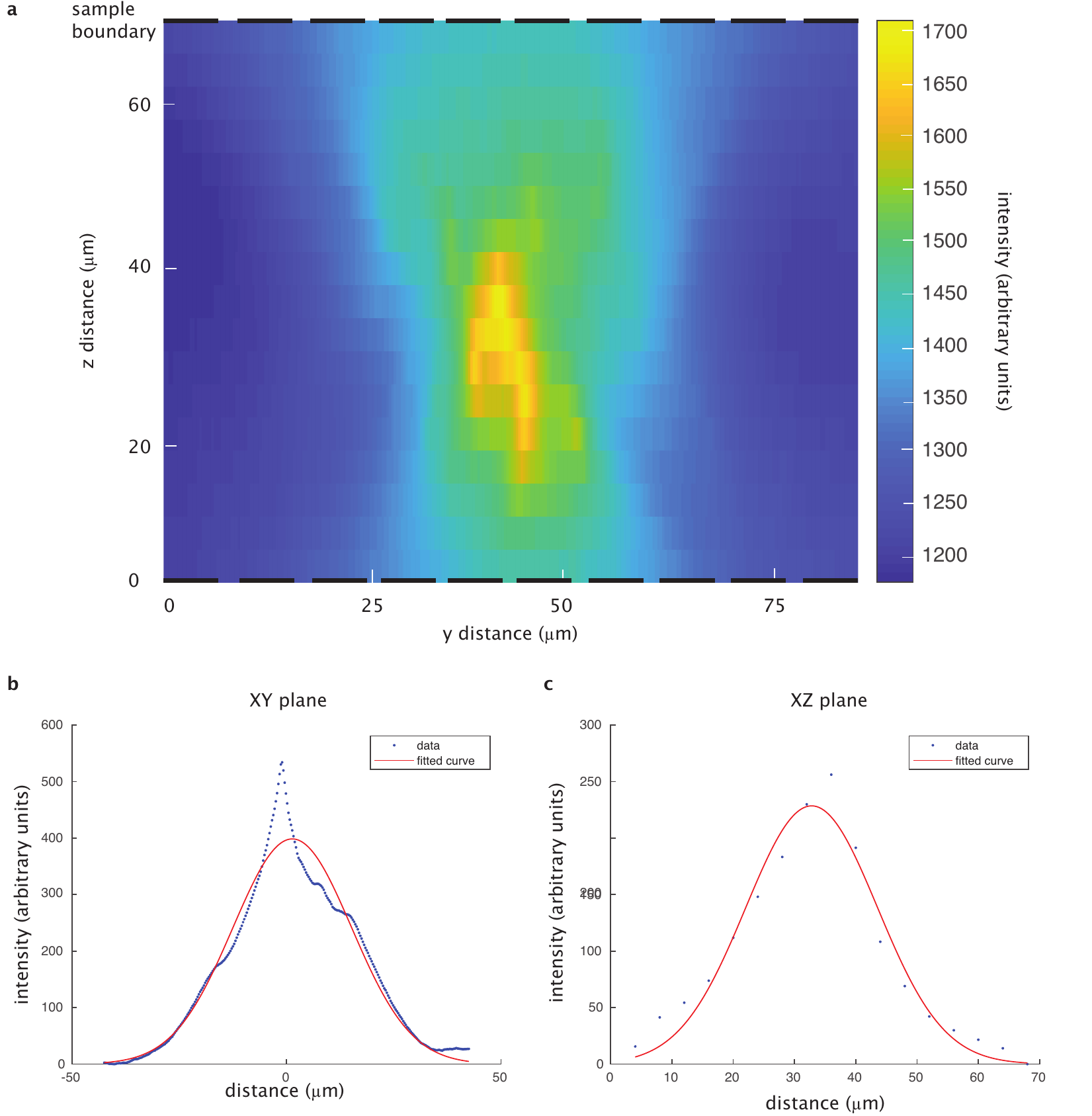}
    \caption{Analysis of microtubule distribution in 3D. \textbf{a}, Heatmap of microtubule distribution in the YZ plane shown in Fig.~\ref{fig:3DAsters}a. Sample boundaries, defined by the coverslips, are denoted by the dashed lines. \textbf{b}, Gaussian fit to the microtubule density in the middle slice of the Y plane. \textbf{c}, Gaussian fit to the microtubule density along the Z plane. }
    \label{fig:3DAnalysis}
\end{figure}

From Z-stack imaging we observe that asters are complex 3D structures (Fig.~\ref{fig:3DAsters}). By analyzing the microtubule density in Z, we find that asters form near the midpoint of the sample plane (Fig.~\ref{fig:3DAnalysis}a). Further, we show that these are symmetric structures by fitting the intensity profiles in the Y plane and Z plane to Gaussians (Fig.~\ref{fig:3DAnalysis}b, c).

\subsection{Comparisons with Similar Systems}\label{dsec:othersys}

\subsubsection{Microtubule Vortices}
The original microtubule-motor system \cite{nedelec_self-organization_1997, surrey_physical_2001} is contractile and shows the formation of microtubule vortices in addition to asters. Microtubule vortices have not been observed in our experiments, however. This is likely due to the substantial differences between the boundary conditions. Experiments where vortices are reported have a channel spacing of 5 $\upmu$m, while our experiments have a channel spacing of $\approx$ 70 $\upmu$m.  A large microtubule vortex forms with a boundary that is 90 um in diameter \cite{nedelec_self-organization_1997}, however, our boundaries are 18 mm x 3 mm. Further our experiments use GMPCPP stabilized microtubules with an average length of 7 $\upmu$m, while the work reporting vortices uses taxol stabilized filaments with length range of $\approx$ 10-100 $\upmu$m. There may also be a significant difference between the acrylamide surface chemistry we use and the agarose chemistry used in the other work.

\subsubsection{Extensile vs. Contractile}
We note that our experimental system results in a contractile network rather than an extensile gel. Recent works have shown that conditions leading to a contractile system require long flexible filaments that are capable of buckling and that undergo limited steric interactions \cite{Belmonte_MSB_2017, Lenz_PRL_2012}. In contrast, the extensile active gel or the active nematic relies on high concentrations of depletion agents to preform bundles of short and stiff filaments, unlike in our system. This suggests that the lack of extensile behavior we observe is unrelated to the optically-dimerizable motors but rather the parameters of the microtubule length and depletion agent. Therefore, there is no inherent limitation in the application of optically-dimerizable motors under extensile conditions.

\subsection{Microscopy Protocol}\label{dsec:dataacquire}
Samples were imaged at 10X (Fig 1c, 1e, 1f, 2d, 4a, 4f, and 4h) or 20X (Fig. 1d, 2b, 3b, 3d, and 3e). For Figures 2e and 2f, the distance span of the merger experiments required us to pool data taken at 10X (500 $\upmu$m and 1000 $\upmu$m separations) and 20X magnifications (175 $\upmu$m, 250 $\upmu$m, and 350 $\upmu$m separations).  For the formation, merging, and movement experiments represented in Figures 1-3, the images of the fluorescent microtubules were acquired every 20 s. For each time point a Z-stack of 5 slices spaced by 10-15 $\upmu$m is taken. For the flow experiments represented in Figure 4, a brightfield image and subsequent fluorescent image were acquired every 4 seconds to observe the tracer particles and microtubules, respectively, without Z-stack imaging.  The increased frame rate was needed to ensure sufficient accuracy of the particle velocimetry. For all experiments, we activated light-induced dimerization in the sample every 20 s with a brief 300 msec flash of 2.4 mW/$\text{mm}^2$ activation light from a $\approx$ 470 nm LED.  The rate of activation was based on the estimated off-rate of the iLID-micro complex \cite{guntas_engineering_2015} of $\approx$ 30 s. The duration of the activation light was empirically determined, by gradually increasing the time in 50 msec increments until we observed the formation of an aster. We note that higher frequencies of activation or longer pulse duration result in contractile activity outside of the light pattern. Typically, one experiment was run per sample. Individual samples were imaged for up to 1 hour.  We placed the time limitations on the sample viewing to minimize effects related to cumulative photobleaching, ATP depletion, and global activity of the light-dimerizable proteins. After several hours, inactivated "dark" regions of the sample begin to show bundling of microtubules.

\subsection{Measuring Aster Spatial Distribution with Image Standard Deviation}\label{dsec:som} 
We interpret the pixel intensity from the images as a measure of the microtubule density. Image standard deviation $\sigma$ is a measure of the width of an intensity-weighted spatial distribution over a region of interest, ROI. We use $\sigma$ to characterize how the the spatial distribution of microtubules evolves in time. For each time point, we first normalize each pixel value $I(x,y)$ by the total pixel intensity summed across the ROI.

\begin{equation}
    I_\text{norm}(x,y) = \frac{I(x,y)}{\sum_{x,y \in \text{ROI}} I(x,y)}
\end{equation}

where $I(x,y)$ is the raw intensity of the pixel at position $(x,y)$ after background subtraction. To find $\sigma$, we define the image variance $\sigma^{2}$ of the intensity-weighted spatial distribution as
 
 \begin{equation}
    \sigma^{2} = \sum_{{x,y} \in \ \text{ROI}} [(x-\Bar{x})^2+(y-\Bar{y})^2]\ I_\text{norm}(x,y),
 \end{equation}

where coordinates $\Bar{x}$ and $\Bar{y}$ are the center of the intensity distribution

\begin{equation}
    \Bar{\mathbf{x}} = \sum_{\mathbf{x} \in \ \text{ROI}} \mathbf{x} \ I(\mathbf{x}).
\end{equation}

\subsection{Characteristic Size of an Aster}\label{dsec:charlength}

\subsubsection{Determining Characteristic Size}
As seen in Fig.~\ref{fig:3DAsters}, the irregularity of aster arm spacings and lengths presents very challenging segmentation issues for the detailed modeling of the microtubule distribution.  Instead, we chose to determine a single characteristic size to represent the spatial distribution of the aster.  First, we perform a maximum projection over the Z-stack for each time point to create a 2D image in the XY plane. To represent the projected 2D image, we chose the image standard deviation approach (Supplementary Information~\ref{dsec:som}) to integrate over the variations in the XY plane. We define the characteristic aster size as the image standard deviation $\sigma$ after $\approx$ 15 min of activation. The characteristic size is used to compare with order-of-magnitude scaling arguments (Supplementary Information~\ref{dsec:asterscaling}).  

\subsection{Image Analysis of Asters}\label{dsec:imageanalysis}

\subsubsection{Image Preparation}
At each time point, each Z-stack of images is summed into a single image in the XY plane. We process each XY image to correct for the non-uniformity in the illumination and background intensity.  We ``flatten'' the non-uniformity of the image with an image intensity profile found in the following process. We take the first frame of the experiment and perform a morphological opening operation with an 80 pixel disk followed by a Gaussian smoothing with a 20-pixel standard deviation. The resulting image is then normalized to its maximum pixel intensity to generate the image intensity profile. Images are flattened by dividing them by the intensity profile. We note that this strategy depends on there being a uniform density of microtubules in the first frame.
 
Once images are flattened, the background is found by taking the last frame of aster formation and calculating the mean intensity of the activated region that is devoid of microtubules. Images are subtracted by this background intensity and thresholded so that any negative values are set to zero.

\subsubsection{Defining the Regions of Interest}
As mentioned in Supplementary Information~\ref{dsec:som}, we determine the image standard deviation over a region of interest (ROI).  For the formation experiments, we define the region of activation as the disk encompassing the aster and the region devoid of microtubules around the aster, after $\approx 15$ min of activation, when formation is complete. To identify this region, we segment the low intensity region around the aster. The low intensity region around the aster is found by subtracting the final frame of aster formation from the first frame of the image acquisition. After subtraction, the void region is the brightest component of the image. We segment this region by performing an intensity and size threshold to create a mask. The aster-shaped hole in the mask is then filled. Using the perimeter of the mask, we calculate the diameter of the disk region of activation.

For analyzing the images for the decay process, we alternatively take a region of interest centered on the aster position (from the last frame of aster formation and found using the intensity weighted center) and proportional to the size of the aster in order to reduce the contribution of microtubules diffusing in from the boundary. This proportionality constant was chosen as the ratio of the ROI diameter to the aster diameter for the aster formed with the 50 $\upmu$m disk, which is 1.63.

\subsection{Reversibility of Aster Formation and Decay}\label{dsec:reversibility}

\begin{figure}[H]
    \centering
    \includegraphics[width=0.55\textwidth]{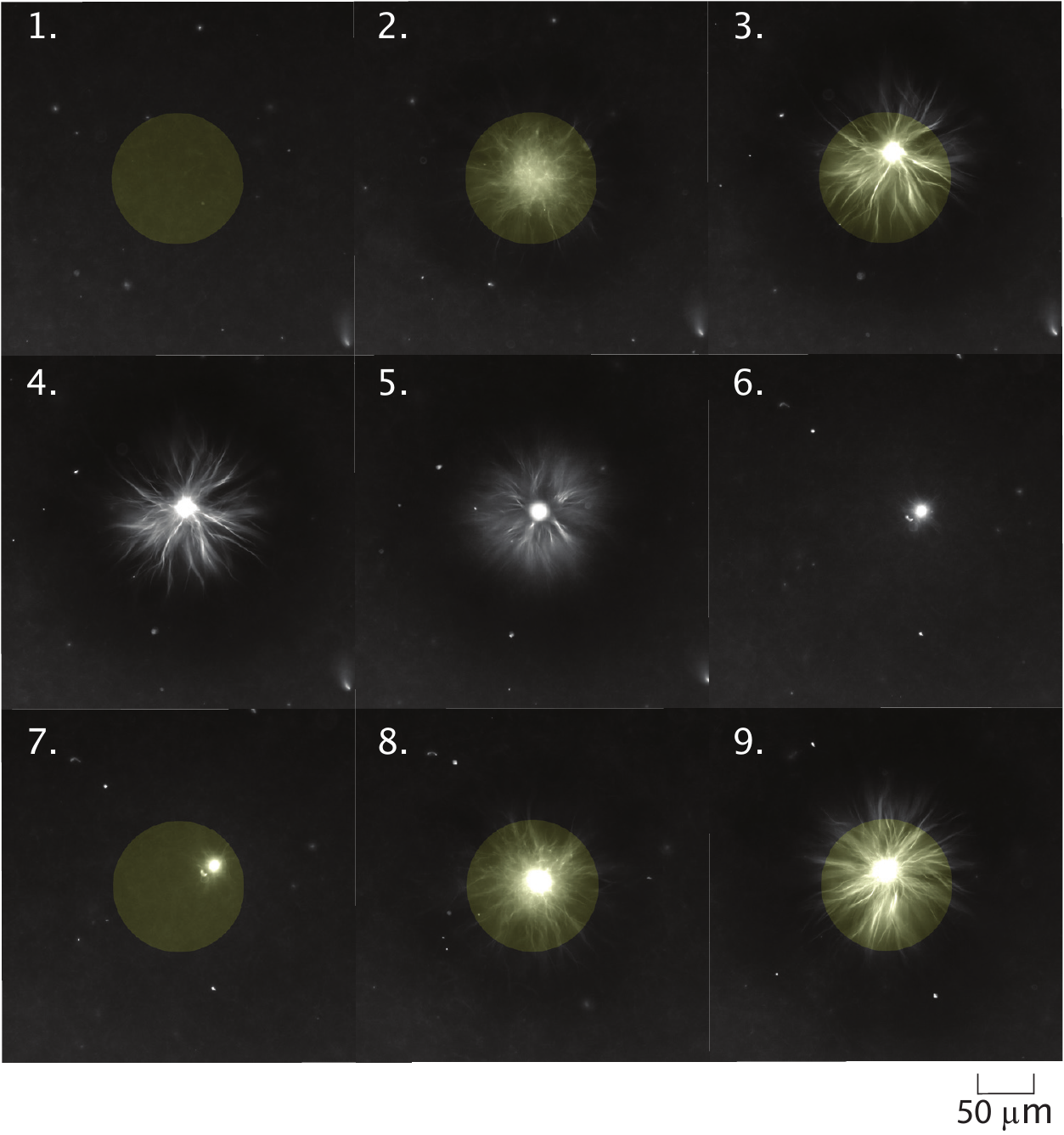}
    \caption{Time series of light induced aster formation, decay, then formation. First formation frames are at time points t = (1) 0, (2) 6.7, and (3) 16.3 min. Aster decay frames are for t = (4) 16.7, (5) 25, and (6) 112.7 min. Second aster assembly frames are t = (7) 113, (8) 120, and (9) 129.3 min }
    \label{fig:form-decay-form}
\end{figure}

To show that aster decay is driven by motors reverting to monomers as opposed to irreversible events such as ATP depletion or protein denaturation, we provide an illustrative experiment of aster formation followed by decay followed again by aster formation. Imaging for this experiment was performed at 20X to increase the spatial resolution. We note that asters do not completely decay, as it is observed in panel 6 of Fig.~\ref{fig:form-decay-form} that the central core of the aster persists.  

\subsection{Speed and Characteristic Time Scales of Formation and Merging}\label{dsec:form}

In order to compare the boundary dependence of our contraction behavior to other contractile networks, we calculate the max speeds and characteristic times of contraction and aster merger as described in~\cite{schuppler_boundaries_2016, Belmonte_MSB_2017,foster2015active}. We first find the characteristic time by fitting a model to our experimental data and then use this value to calculate the maximum speed. As in~\cite{schuppler_boundaries_2016}, we fit to a model of a critically damped harmonic oscillator,

\begin{equation}\label{eq:cdho}
    L(t) = L_\text{fin} + ( L_\text{init} -  L_\text{fin})\left(1 + \frac{t}{\tau}\right)e^\frac{-t}{\tau} ,
\end{equation}

\begin{figure}[H]
    \centering
    \includegraphics[width=.85\textwidth]{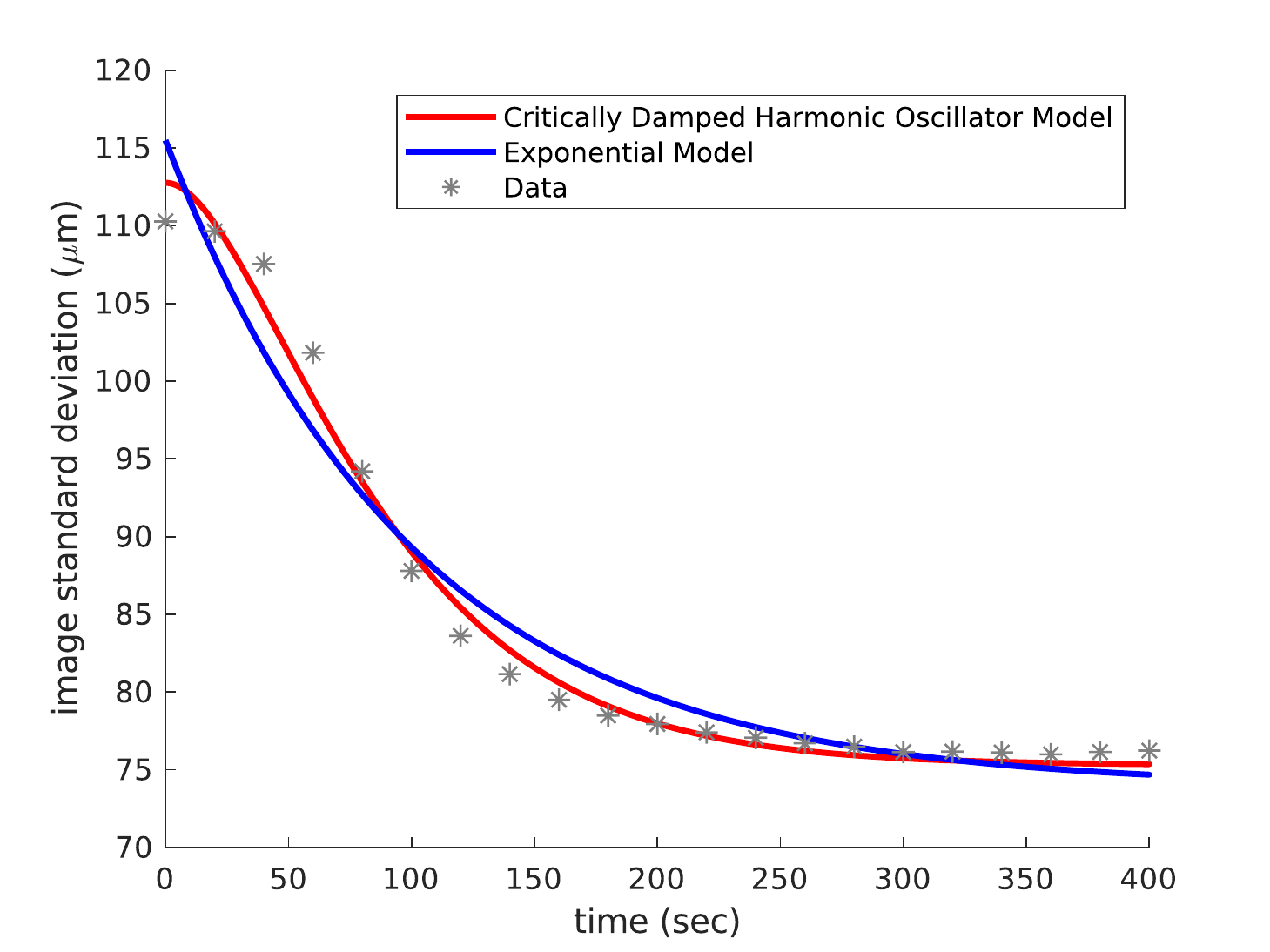}
    \caption{A comparison of model fittings for a contracting aster experiment.}
    \label{fig:expvsho}
\end{figure}

where $L_\text{init}$ is the initial size of the network, $L_\text{fin}$ is the final network size, and $\tau$ is the characteristic time of contraction. This model was developed to describe a contractile actomyosin gel, which shares similar dynamics with our own system. We apply this fit on time points after the initial lag phase, which was empirically determined to end at one minute. While we tried fitting to an exponential function, we found that the harmonic oscillator model was more robust across excitation length scales (Fig.~\ref{fig:expvsho}).

\begin{figure}[H]
    \centering
    \includegraphics[width=.85\textwidth]{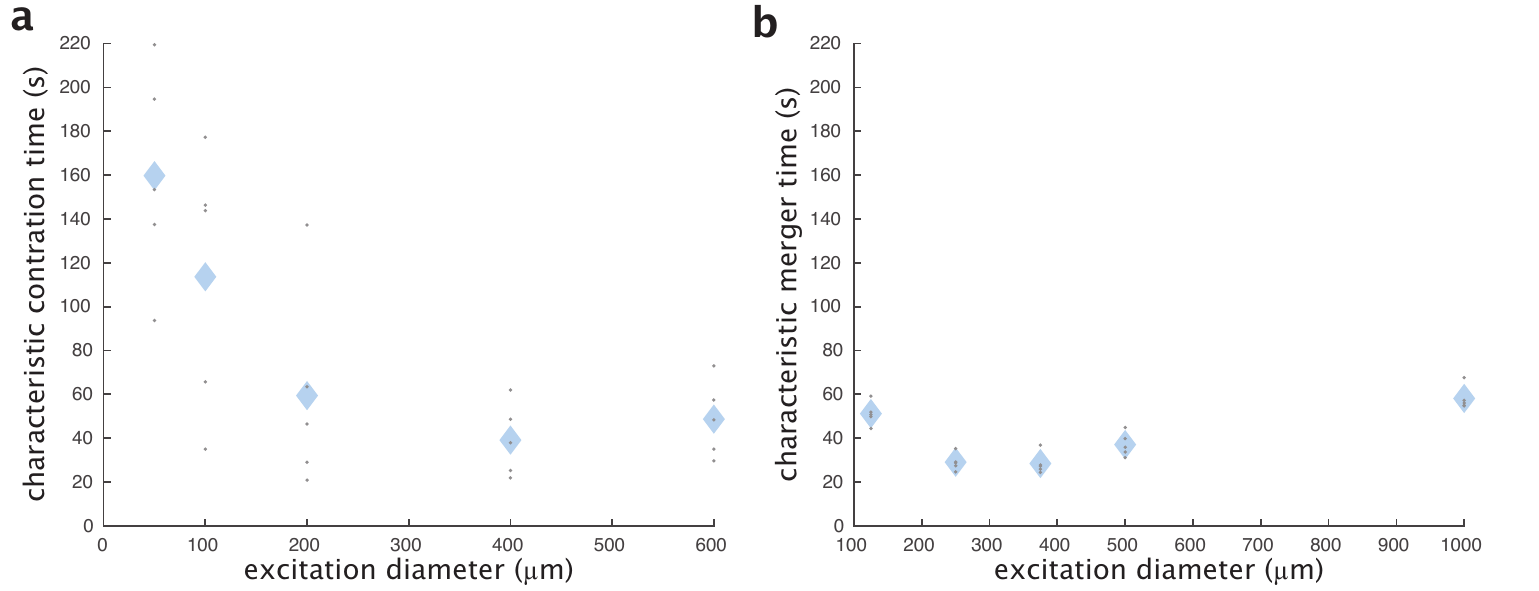}
    \caption{Characteristic times for contraction and merger as functions of activation length scales. \textbf{a}, Characteristic time for aster formation as a function of the excitation diameter. \textbf{b}, Characteristic time for aster merging as a function of the initial distance between asters.}
    \label{fig:chartime}
\end{figure}

 We find that the characteristic times show a general lack of sensitivity to system size for our range of lengths (Fig.~\ref{fig:chartime}), similar to \cite{foster2015active}. The characteristic time is roughly 1 to 2 minutes, comparable to the times reported in \cite{foster2015active}.
 
 We calculate the maximum speed of contraction or merger, $v_\text{max} =  - \frac{dL(t_\text{max})}{dt}$, by finding the time $t = t_\text{max}$ that satisfies $ \frac{d^2L(t_{max})}{dt^2} = 0$. First we calculate the second derivative of Eq.~\ref{eq:cdho},
 
 \begin{equation}
     \frac{d^2L(t)}{dt^2} = \frac{(L_\text{init} - L_\text{fin})(t-\tau)}{\tau^3} e^{\frac{-t}{\tau}} .
 \end{equation}
 
 Based on this equation, it is apparent that the maximum speed occurs at $t_\text{max} = \tau$. The maximum speed is then defined as, $v_\text{max} = - \frac{dL(\tau)}{dt}$. We calculate $-\frac{dL(t)}{dt}$ by taking the first derivative of Eq.~\ref{eq:cdho},

 \begin{equation}
     \frac{dL(t)}{dt} = \frac{t(L_\text{init} - L_\text{fin})}{\tau^2} e^\frac{-t}{\tau},
 \end{equation}

then set $t = \tau$ to find the maximum speed,

\begin{equation}
    v_\text{max} = \frac{dL(\tau)}{dt} = \frac{L_\text{init} - L_\text{fin}}{e \tau} .
\end{equation}

This $v_\text{max}$ is the reported contraction or merger speed. 

\subsection{Comparison to Light Activated Actomyosin Networks}\label{dsec:NetworkComp} 
A system that shows some similar behavior to ours is the light activated actomyosin network in \cite{schuppler_boundaries_2016}. Here, we note the similarities and differences between the two systems. In the actomyosin network, the actin filaments are globally and permanently crosslinked by the myosin motors in both the dark and the light. In the light, motors are permanently activated. Light patterns generate a localized contraction of the global actomyosin network.  Since the contracting region is still linked to the rest of the actomyosin network, deformations are propagated throughout the entire network.  

In contrast, our system starts with unlinked microtubule filaments. Light patterns activate linkages of motors to create a localized contractile network with a free boundary. Thus, there are no connections to an external network, unlike the actomyosin system. Further, the reversibility of these links allows the networks to remodel and to resolve after contraction. 

A key similarity between the two systems is the observation that contraction speed increases linearly with the size of the excitation region. A recent theoretical treatment \cite{Belmonte_MSB_2017} provides a generic model for this observation.  Their results in Box 1 Panel C predict a linear scaling of contraction speed versus size for 1D, 2D, and 3D networks. For a 1D network, the contraction speed $\frac{dL}{dt}$ is related to the length $L$ of the network by the contractility constant $\chi$ as

\begin{equation}
\frac{dL}{dt} \approx \chi L .
\end{equation}

\subsection{Analysis of Aster Decay}\label{dsec:decay}

When the activation light is removed, the iLID-micro dimer begins to disassociate, leading to un-crosslinked microtubules. The original work where iLID is designed and characterized show that the formation and reversion half-lives of individual iLID-micro heterodimers are on the order of 30 seconds  \cite{guntas_engineering_2015}. 

Our empirical determination that sharp localization of contractile forces within the light pattern requires pulsing the light pattern every 20 seconds (Supplementary Information~\ref{dsec:dataacquire}), in addition to the characterization of other iLID and LOV domain based systems \cite{Tas_LightGlide_2018, nakamura2018intracellular,johnson2017spatiotemporal,yumerefendi2016light}, supports the notion that the reversion rate of kinesin-fused iLID proteins is similarly on the tens of seconds time scale. We note that the motor density has been predicted and observed to increase exponentially towards the aster center \cite{nedelec_dynamic_2001}. We therefore expect the central region of the aster to decay more slowly than an individual motor link. This may explain why asters appear to decay on the order of tens of minutes (Fig. 1c), rather than tens of seconds. 

For an ideal 2D Gaussian spatial distribution of diffusing particles starting with a finite radius of $w$, we expect 
\begin{equation}
    p(r,t) = \frac{1}{\pi( 4D t + w^2)}e^{-r^2/(4Dt+w^2)},
\end{equation}
where $D$ is the diffusion coefficient.

The variance $\sigma_\text{Gauss}^2$ of this distribution as a function of time $t$ is given by 

\begin{equation}
    \sigma_\text{Gauss}^2(t) = 4D t +  w^2.
\end{equation}

The variance $\sigma_\text{Gauss}^2$ increases linearly with $t$ with a slope of 4$D$.

\begin{figure}[H]
    \centering
    \includegraphics[width=0.55\textwidth]{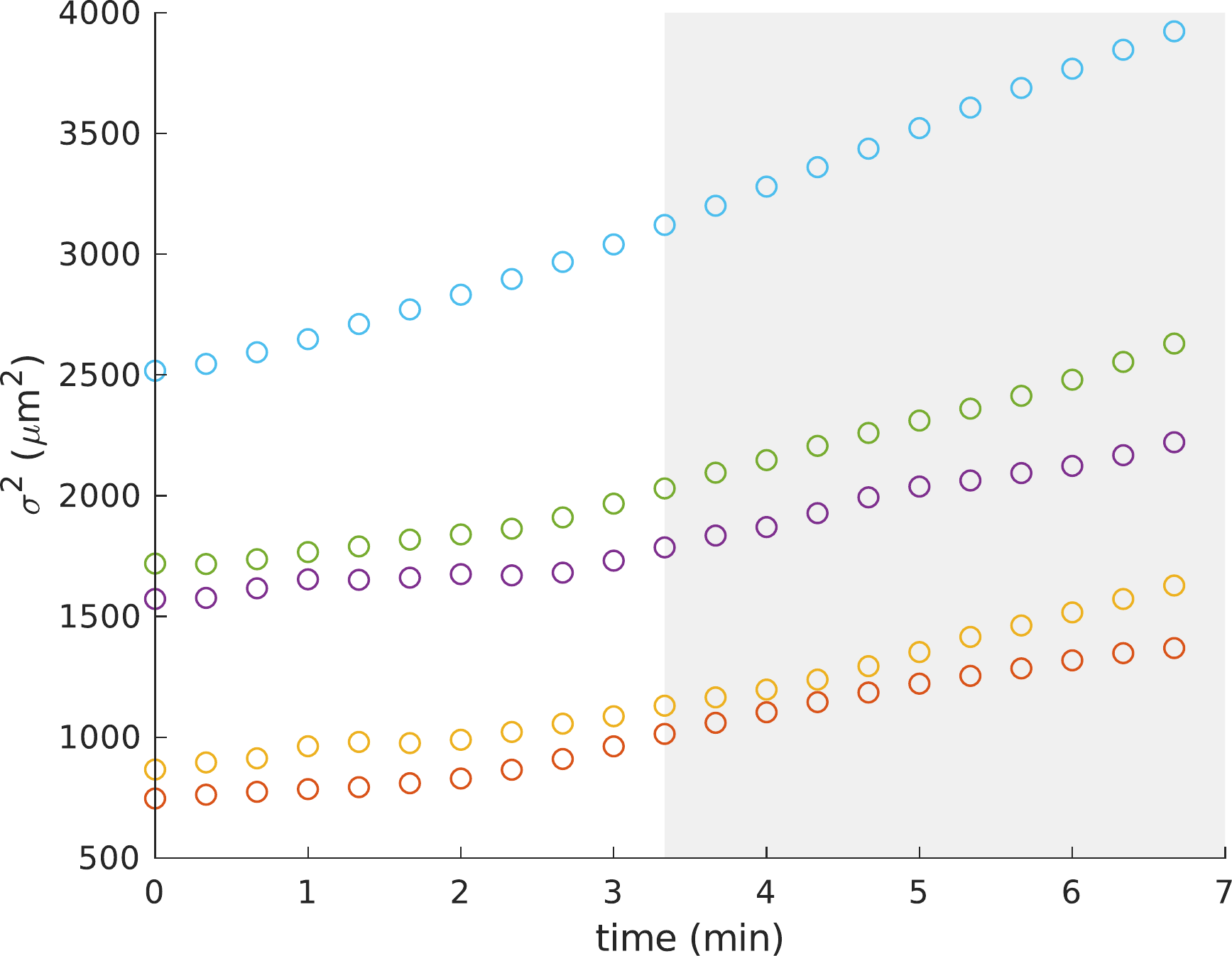}
    \caption{Plot of mean variance of image intensity as a function of time for different initial aster sizes. The shaded region is treated as part of the linear regime. The measure of time is relative to the beginning of aster decay.}
    \label{fig:decay-variance}
\end{figure}

We characterize the aster decay process by measuring the image variance $\sigma^2$, as a function of time, as described in (SI.~\ref{dsec:som}). Images are first processed as described in (SI.~\ref{dsec:imageanalysis}). Although our spatial distributions are not strictly Gaussian, we observe that for our data that $\sigma^2$ increases linearly with $t$ (Fig.~\ref{fig:decay-variance}), which suggests that the decay process is described by the diffusion of unbound microtubules. By analogy to the 2D ideal Gaussian case, we calculate an effective diffusion coefficient of our distributions by a linear fit of $\sigma^2$ versus time and finding the diffusion coefficient from the slope. This gives us a diffusion coefficient in units of $\upmu$m$^2$/s.

We find the diffusion coefficient by applying a linear fit to time points that occur after 200 seconds.

\subsection{Diffusion Coefficient of a Microtubule} \label{dsec:drag}
We estimate the diffusion coefficient for a single microtubule to compare with the effective diffusion coefficient we estimate for aster decay. The diffusion coefficient $D$ for an object in liquid media can be calculated from the drag coefficient $\gamma$
\begin{equation}
    D = \frac{k_{B}T}{\gamma},
\end{equation}

where $k_{B}$ is the Boltzmann constant and T is the temperature, for which we use 298 K. We model a microtubule as a 7 $\upmu$m long cylinder (SI.~\ref{msec:polymerization}) with a radius of 12.5 nm. The drag coefficients for a cylinder have been found previously \cite{Tirado1979_DiffCyl} for motion either parallel $\gamma_{\parallel}$ or perpendicular $\gamma_{\perp}$ to the long axis of the cylinder

\begin{equation}
  \begin{aligned}
    \gamma_{\parallel} = \frac{2 \pi \eta L}{\ln(L/2r) - 0.20}, \\
    \gamma_{\perp} = \frac{4 \pi \eta L}{\ln(L/2r) + 0.84}.
    \end{aligned}
\end{equation}

Here, $L$ is the length of the cylinder, $r$ is its radius, and $\eta$ is the viscosity of the fluid, which we estimate to be $2 \times 10^{-3}$ $\text{Pa} \cdot \text{s}$ (SI~\ref{dsec:viscosity}). Using the parameters detailed above, we calculate $D_{\parallel}$ = 0.3 $\upmu\text{m}^2/$s and $D_{\perp}$ = 0.2 $\upmu\text{m}^2/$s. We assume that the larger diffusion coefficient dominates and thus use $D_{\parallel}$, the longitudinal diffusion coefficient, as the diffusion coefficient for a single microtubule in Fig.~1e.

\subsection{Scaling Arguments for Aster Size and Comparison to Data} \label{dsec:asterscaling}  
\begin{figure}[H]
\centering
\includegraphics[width=0.9\textwidth]{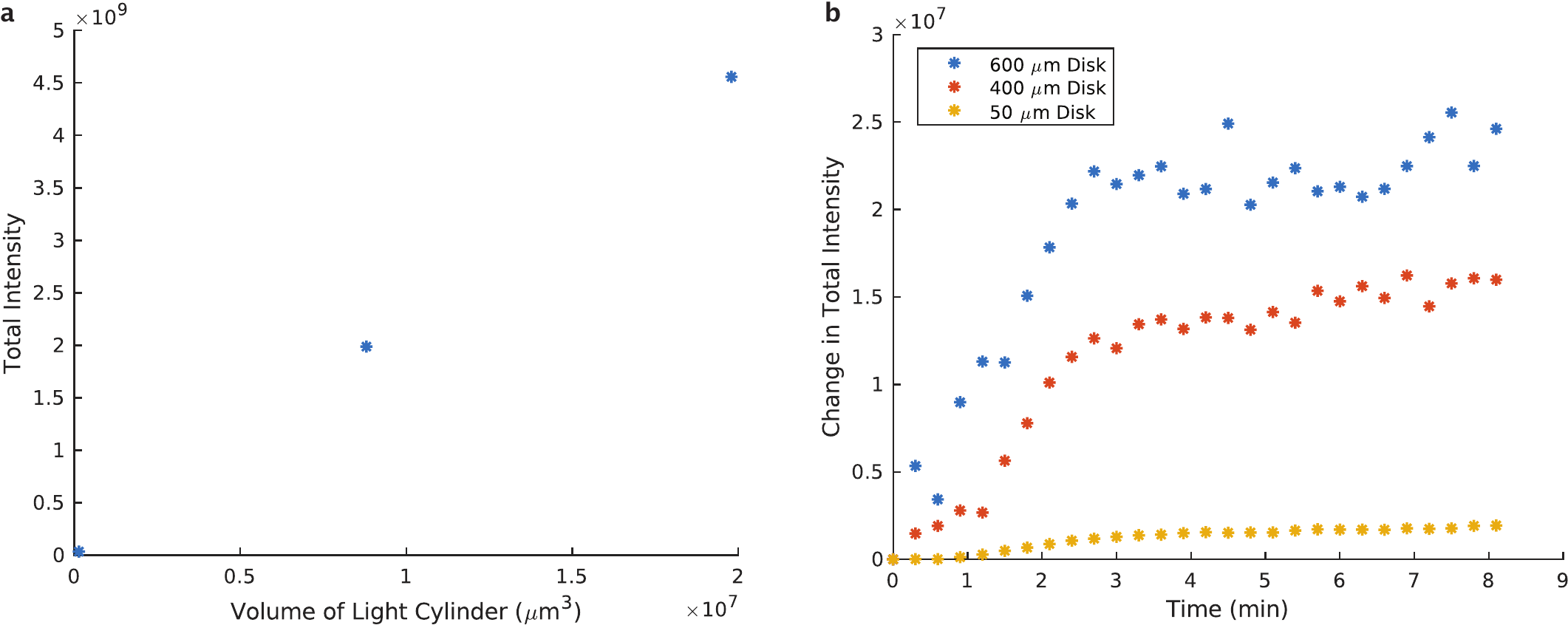}
\caption{\label{fig:int}Measuring the conservation of labeled fluorescent microtubules in the excitation region during aster formation. \textbf{a}, Total intensity of excitation region as a function of volume of light cylinder averaged during aster formation. Measurements are for light disks with diameters 50, 400, and 600 $\upmu$m. \textbf{b}, Change in total intensity inside of the excitation region as a function of time}
\end{figure}

We consider how the total number of microtubules in an aster relates to the volume of the projected light pattern. We are projecting a disk pattern of light on the sample from below. The channel is a constant height, $z \approx 70 \, \upmu$m. We therefore treat the light excitation volume as a cylinder $V_\text{light} = \frac{1}{4} \pi z d_\text{light}^2$ where $d_\text{light}$ is the diameter of the excitation disk. If we look at experimental data, we see evidence of a linear relationship between the light volume and the number of microtubules that are present during aster formation (Fig.~\ref{fig:int}a). The implication of this observation is that the density $\rho$ of microtubules is uniform. Furthermore, we see that after the initial contraction event, the total integrated fluorescence of the excited region remains constant (Fig.~\ref{fig:int}b), indicating that the total number of microtubules $N$ is constant during aster formation.

Based on these observations we assume the number of microtubules $N$ in the aster is given by
\begin{equation}
    N \approx \rho V_\text{light}.
\end{equation}

From Supplementary Information ~\ref{dsec:3D}, we observe that asters have a roughly spheroidal symmetry.  For an order-of-magnitude estimate of how aster size scales with the volume of light,  we assume the characteristic length of the aster $L_\text{aster}$ is given by the diameter of an effective sphere which scales with microtubule number as
\begin{equation}
    L_\text{aster} \propto N^{1/3}.
\end{equation}
and thus 
\begin{equation}
    L_\text{aster} \propto V_\text{light}^{1/3}.
\end{equation}

As noted above, the volume defined by the activation light is a cylinder, then
\begin{equation}
    V_\text{light} \propto d_\text{disk}^2.
\end{equation}
From these last two equations, we arrive at the scaling relationship between aster size and excitation disk size
\begin{equation}
    L_\text{aster} \propto d_\text{disk}^{2/3}.
\end{equation}

We made a power law fit with a fixed exponent of 2/3 to the data shown in Fig.~1f. Though we cannot strictly rule out other exponents,  we show the fit to demonstrate that the scaling argument determined exponent is at least consistent with the data.     

\subsection{Tracking of Moving Aster}\label{dsec:astertrack}
For each time point, we sum over the z-stack to form a single image. The image is then passed through a morphological top-hat filter with a structure element of a 100 pixel disk to ``flatten'' non-uniformities in the illumination. The image is then projected into a 1D intensity profile. We project onto the x-axis by summing along the line that passes through the center of the excitation disk with a 100 pixel window in y. Aster centers are then found at each frame by fitting the intensity profiles to Gaussian functions.

For 2D tracking, the movement of the aster is found by comparing the centroid of the aster in each frame. The raw images are processed using a Gaussian filter with a standard deviation of $1$ pixel, followed by thresholding to eliminate the background noise.

\subsection{Effective Potential of a Moving Aster}\label{dsec:movepotential}
\begin{figure}[H]
    \centering
    \includegraphics[width=.55\textwidth]{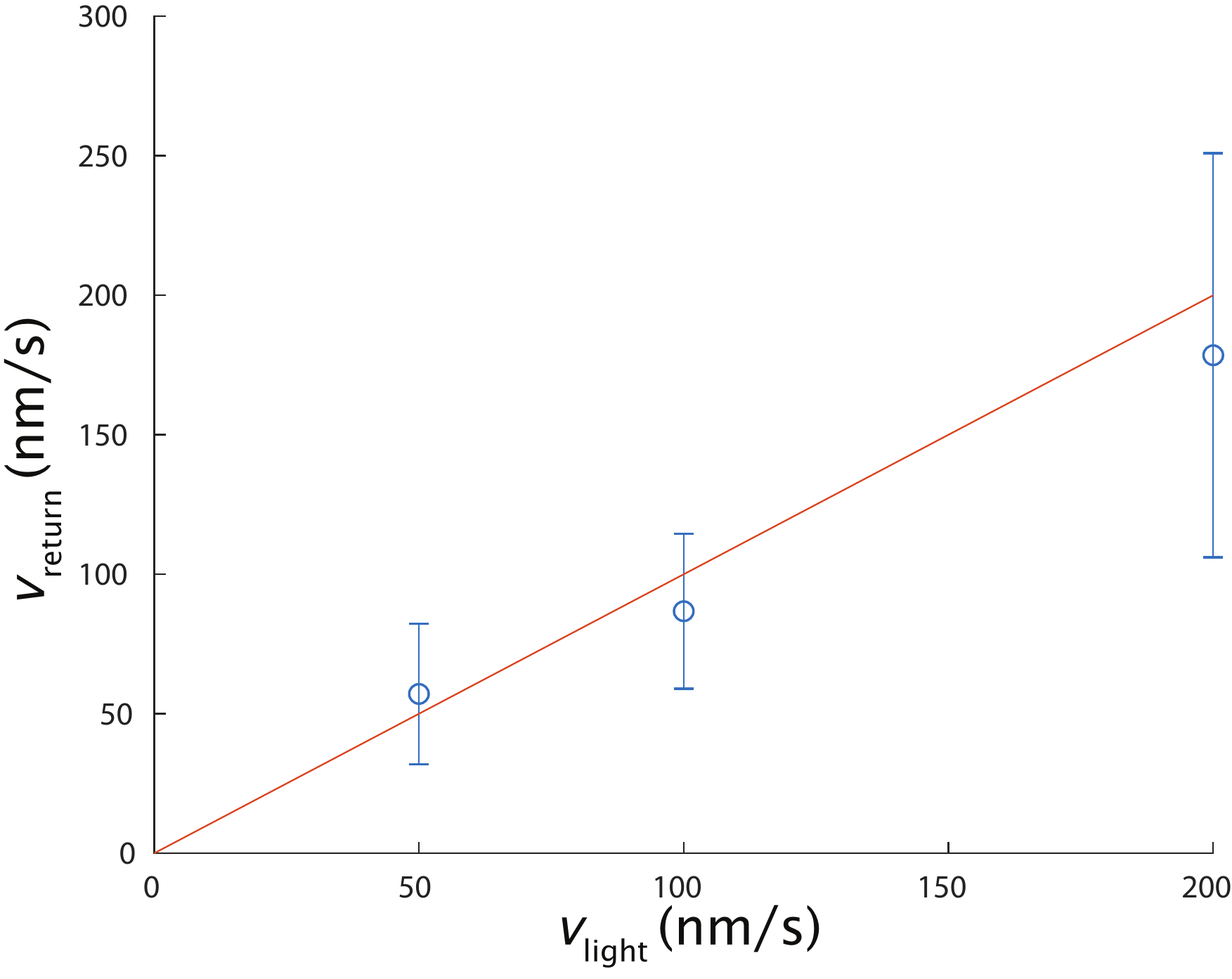}
    \caption{The speed at which an aster returns to the center of the light pattern once the pattern stops moving. Red line is a plot of y = x.}
    \label{fig:snapback}
\end{figure}

When the light pattern moves, we observe the aster appears to be pulled in tow behind the light pattern, perhaps by the aster arms or newly-formed microtubule bundles in the light pattern.  Further, when the light pattern stops moving at speed $v_\text{light}$, we observe the aster immediately returns to the center of the light pattern at speed $v_\text{return}$. From the Fig.~\ref{fig:snapback}, we see that 
\begin{equation}
    v_\text{return} \approx v_\text{light}.
\end{equation}

This is the behavior expected for an object under the influence of a potential at low-Reynolds-number, where the aster has negligible momentum and the forces are essentially instantaneous. These observations support the notion that a moving aster can be modeled as being in an effective potential. First, we model the observed behavior with a generic potential without any assumption of the mechanistic cause of the potential and then numerically compare these results to the estimated optical tweezer effects of the excitation light pattern.

We estimate the potential and the forces acting on a moving aster from the viscous drag of the background fluid, in an analogous way to how this is done for objects trapped in an optical tweezer \cite{svoboda_tweezer1994}.  If we assume the aster is a spherical object of radius $a$ and is moving with speed $v_\text{light}$, it will experience a viscous drag force $F_\text{drag}$ : 
\begin{equation}
 F_\text{drag} = 6 \pi \eta a v_\text{light},
\end{equation}
where $\eta$ is the fluid viscosity.  $F_\text{drag}$ is equal to the force $F_\text{pull}$ that is pulling the aster towards the light pattern. From the results of Fig. 2c, we note the observed distance shift $\ell$ of the aster from the center of the moving light pattern is roughly linear with excitation disk movement speed $v_\text{light}$. The linearity of $\ell$ versus $v_\text{light}$ implies that $F_\text{pull}$ acts like a spring:
\begin{equation}
 F_\text{pull} \approx k_\text{spring} \ell,
\end{equation}
where $k_\text{spring}$ is the spring constant.  Setting these two forces equal gives a spring constant of
\begin{equation}
 k_\text{spring} \approx \frac{6 \pi \eta a v_\text{light}}{\ell}.  
\end{equation}
The effective potential $U_\text{pull}$ for this force is 

\begin{equation}
    U_\text{pull} = \frac{1}{2} k_\text{spring} \ell^2.
\end{equation}

The aster in Fig. 2c is  $\approx$ 25 $\upmu$m in diameter. Assuming that $\eta \approx 2 \times 10^{-3}$ $\text{Pa} \cdot \text{s}$ (SI~\ref{dsec:viscosity}), we find that $k_\text{spring} \approx 3 \times 10^{-15}\, \text{N}/\upmu \text{m}$. For the maximum observed displacement of $\ell$ $\approx$ 30 $\upmu$m, the energy stored in the potential, or equivalently, the work done by the system to return the aster back to the center of the light pattern is  $\approx$ 300 $k_{B}T$.  

The spring constant of an optical tweezer trapping polystyrene spheres is   $\approx$ 1$\times 10^{-9}\, \text{N}/\upmu \text{m}$ for a  $\approx$ 1000 mW laser beam focused to  $\approx$ 1 $\upmu$m diameter \cite{Mahamdeh2011_tweezer}. Accounting for light intensity, we estimate the spring constant to be  $\approx$ 1 $\times 10^{-12}\, \text{N}/\upmu \text{m}$ per mW/$\text{$\upmu$m}^2$. In comparison, our light pattern has intensity of 2.4 mW/$\text{cm}^2$. The light is on only for 0.3 sec every 20 sec (SI~\ref{dsec:dataacquire}), giving a time averaged intensity of 0.036 mW/$\text{cm}^2$.  The estimated upper bound spring constant from the light pattern due to optical tweezing effects is  $\approx 3.6 \times 10^{-22}\, \text{N}/\upmu \text{m}$, roughly a factor of $10^7$ weaker than the spring constant we observe. Further, we note that it is a generous assumption that a microtubule aster is refractile as a polystyrene sphere. Given the unlikelihood of optical tweezing being related to the potential we observe, we attribute the effective potential other effects such as the remodeling of the microtubule field.

\subsection{Mechanism and Stability of a Moving Aster}\label{dsec:astermech}

\begin{figure}[H]
    \centering
    \includegraphics[width=.85\textwidth]{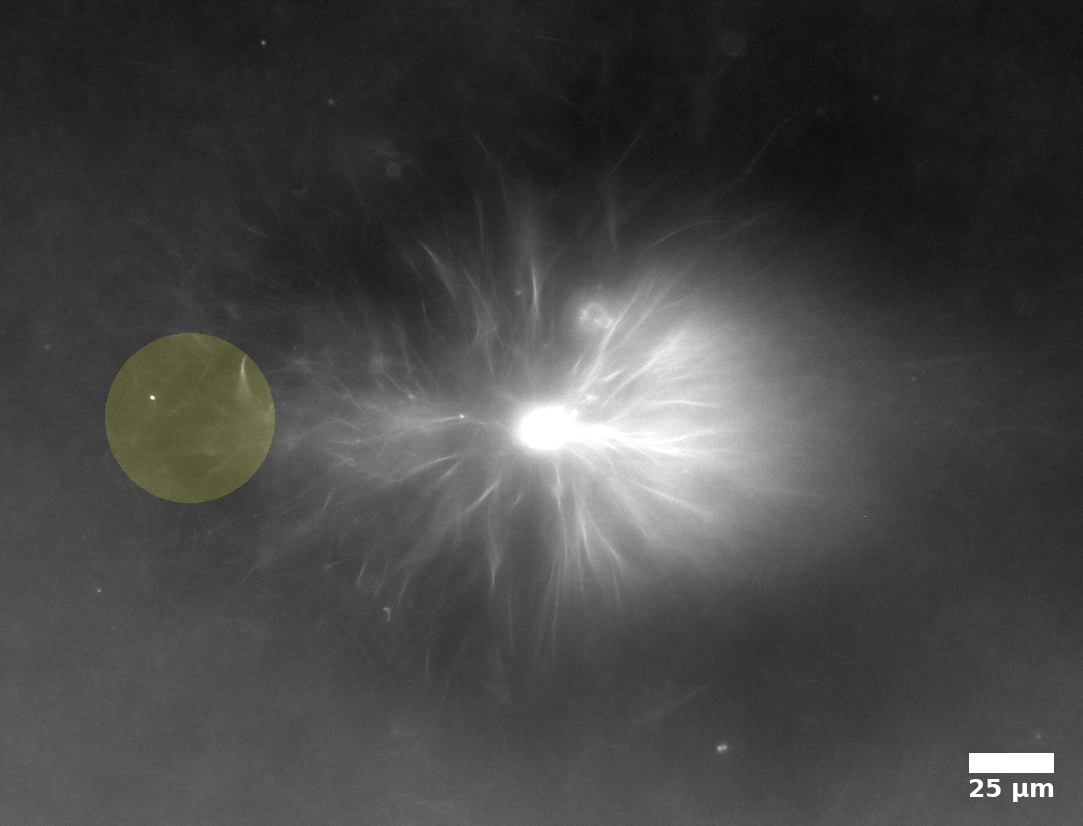}
    \caption{Aster following a 50 $\upmu$m disk moving at 200 nm/s from right to left. Image is integrated across z.}
    \label{fig:largemovingaster}
\end{figure}

While the molecular details of aster movement remains a topic of future study, there are mesoscopic phenomena that we observe. When the light pattern activates a region adjacent to the aster, microtubule bundles form.  As the light pattern moves, a stream of bundles spans from the light pattern towards the aster. This behavior can be most clearly seen at the highest stage speeds of 200 nm/s and with larger disk sizes (Fig.~\ref{fig:largemovingaster}). 

The stream of bundles appears to pull against the arms of the aster towards a new contractile center.  

\begin{figure}[H]
    \centering
    \includegraphics[width=.7\textwidth]{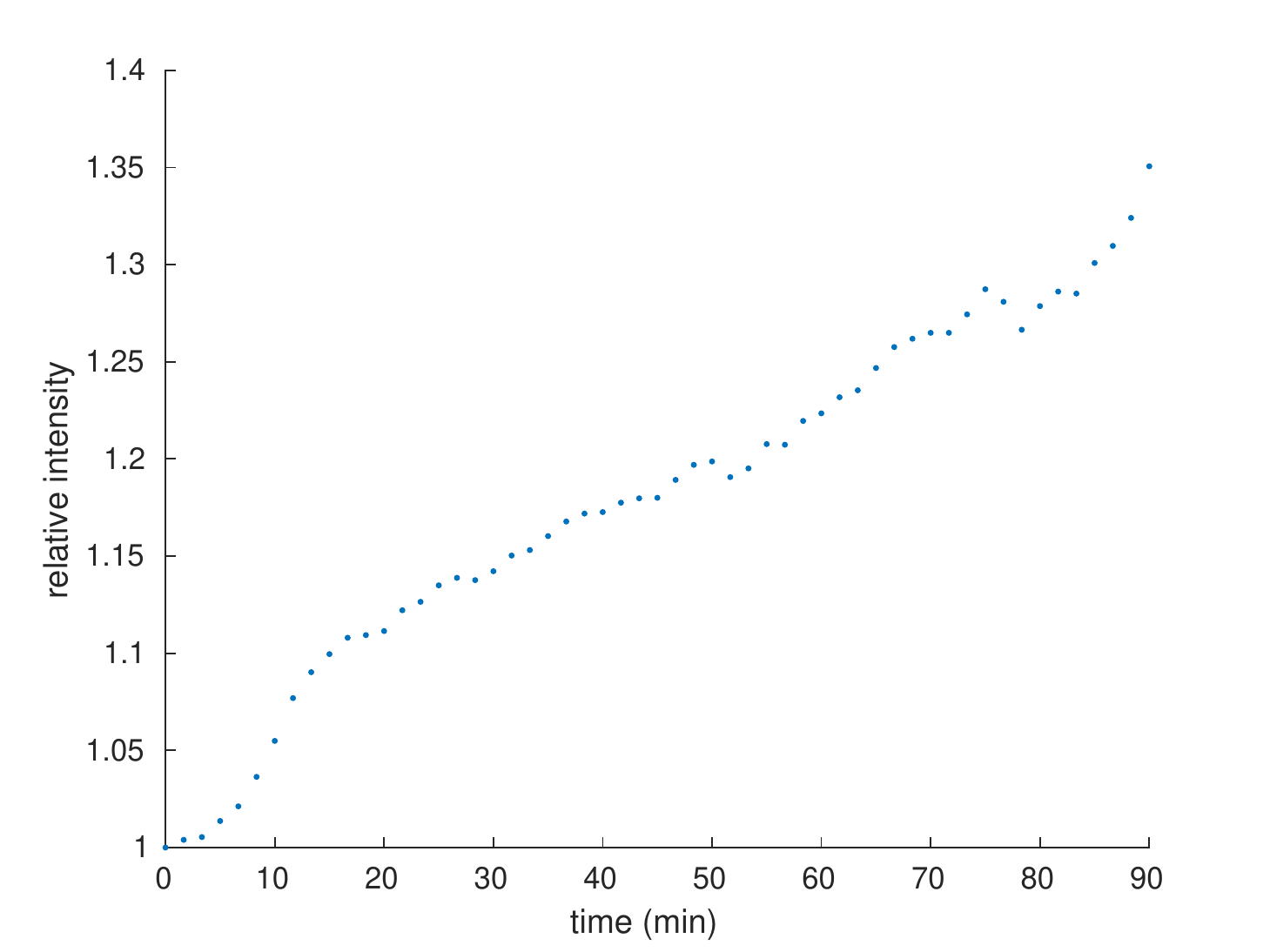}
    \caption{Intensity of an aster for a light pattern moving at 200 nm/s. The y-value is normalized to the intensity at t = 0. Intensity is measured for an ROI with a fixed diameter and tracks with the aster center.}
    \label{fig:moveint}
\end{figure}

During aster movement, we observe a cloud of unbundled microtubules are left in the wake of a moving aster, indicating that there is a decay process occurring. At the same time, however, we also observe microtubules are incorporated into the aster, as demonstrated by the increase in the aster intensity over time (Fig.~\ref{fig:moveint}), which starts to occur after a few minutes. The increase in intensity also indicates that the incorporation rate is greater than the aster decay rate. We speculate that the newly added microtubules deliver linked motor proteins that maintain some of the bonds between filaments, allowing the aster to persist outside of the light pattern.

\subsection{Single Motor Velocity Determination from Gliding Assay}\label{dsec:glidingspeed}
Gliding assay images were acquired every second with total internal reflection fluorescence (TIRF) microscopy. Motor speeds were determined by tracking individual microtubules. Single microtubules were identified by edge detection followed by size thresholding to remove small particles on the glass and large objects that are overlaying microtubules. The centroid of each object is identified and paired with the nearest-neighbor in the next frame. The Euclidian distance between the paired centroids is calculated and used to determine the microtubule velocity. The mean motor speed was determined from the mean frame-by-frame velocities (excluding those less than 75 nm/s, which is our typical sample drift). 

\begin{figure}[H]
    \centering
    \includegraphics[width=.7\textwidth]{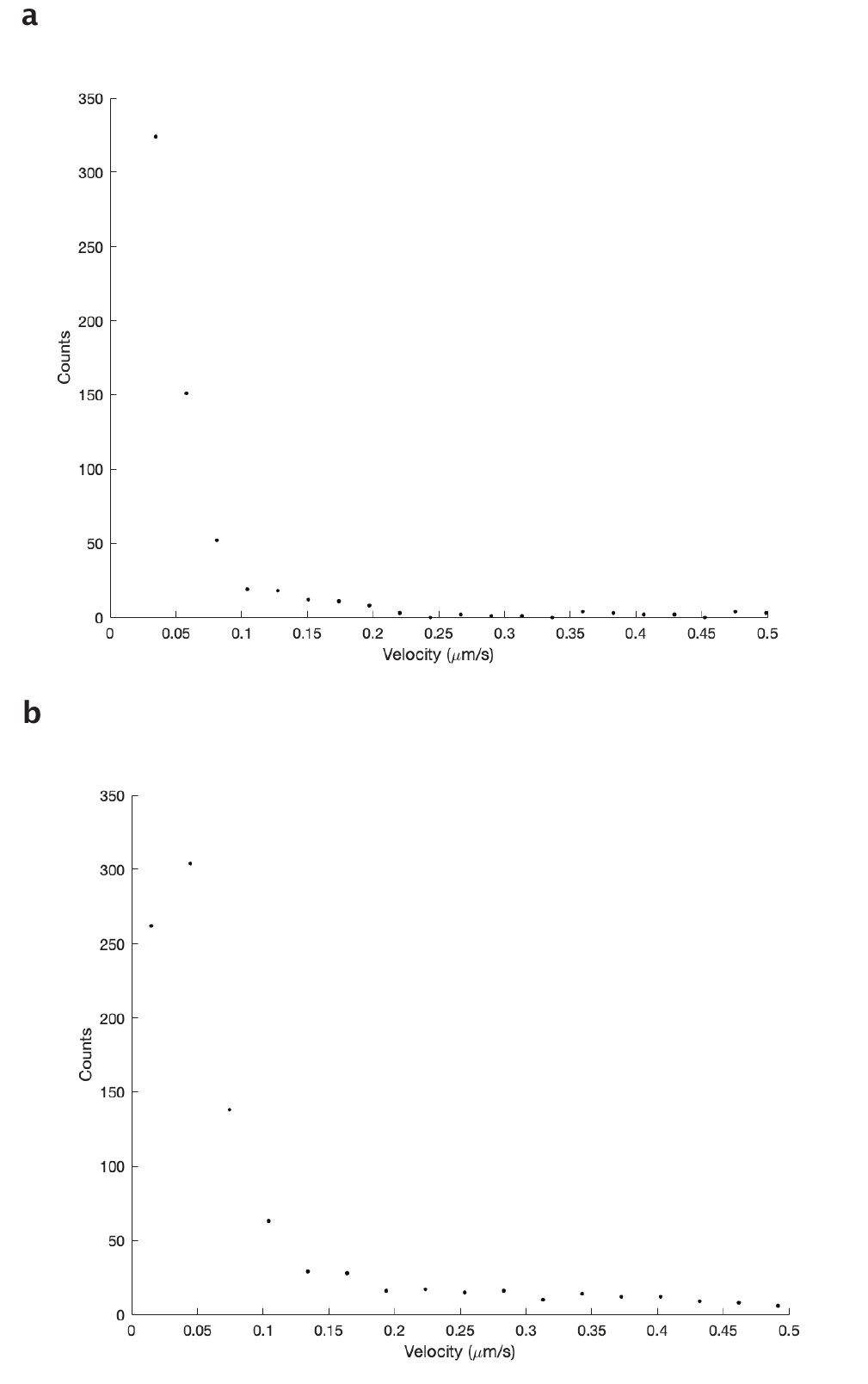}
    \caption{Velocity distribution of gliding microtubules. \textbf{a}, Binned velocities for K401-iLID motors, the mean of the data is 230 nm/s with a standard deviation of 200 nm/s. \textbf{b}, Binned velocities for K401-micro motors, the mean of the data is 300 nm/s with a standard deviation of 250 nm/s.}
    \label{fig:motorspeeds}
\end{figure}

\subsection{Minimum Size Limits of Structures}\label{dsec:minsize}

\begin{figure}[H]
    \centering
    \includegraphics[width=.95\textwidth]{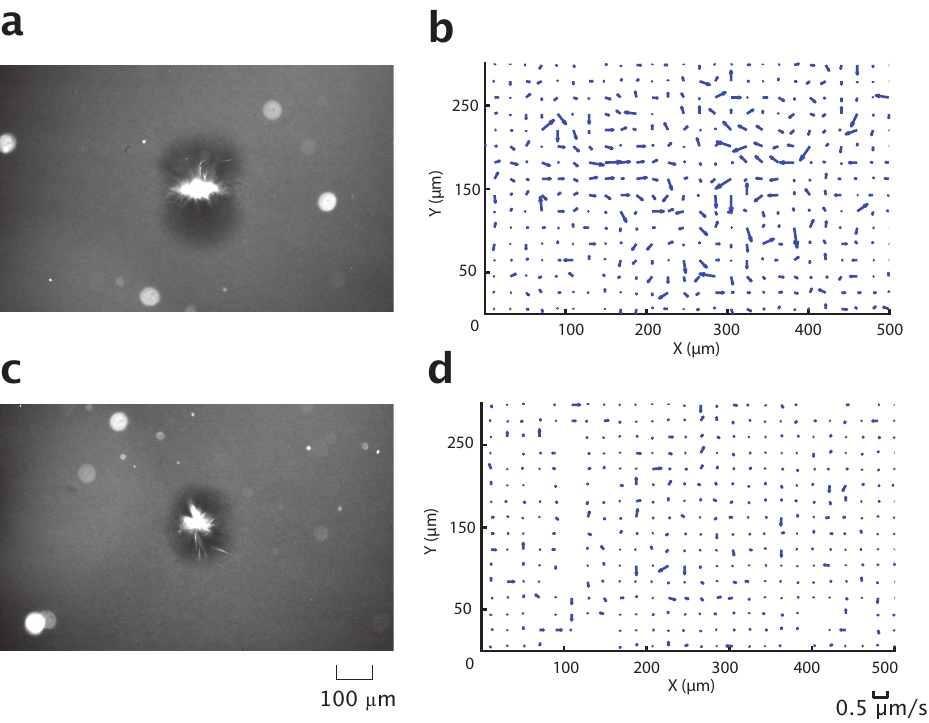}
    \caption{Minimum length experiment for a L x 20 $\upmu$m excitation pattern. \textbf{a}, Fluorescent microtubule channel for L $= 175$ $\upmu$m. \textbf{b}, Corresponding flow field to (\textbf{a}). \textbf{c}, Fluorescent microtubule channel for L $= 87.5$ $\upmu$m. \textbf{d}, Corresponding flow field to (\textbf{c}).}
    \label{fig:minlength}
\end{figure}

Here we explore the minimum feature sizes that we can generate. To test the limits for flow generation, we vary the length and height of the excitation bar. We observe that the minimum excitation bar length that is able to generate flows is between 87.5-175 $\upmu$m Fig.~\ref{fig:minlength}, which corresponds to a microtubule network of $\approx 100$ x 30 $\upmu$m. We note that this length is similar to the bundle buckling length observed in Fig.~4b. We speculate that the limits of the minimum length pattern for generating flow may be related to this buckling length scale. 

\begin{figure}[H]
    \centering
    \includegraphics[width=.95\textwidth]{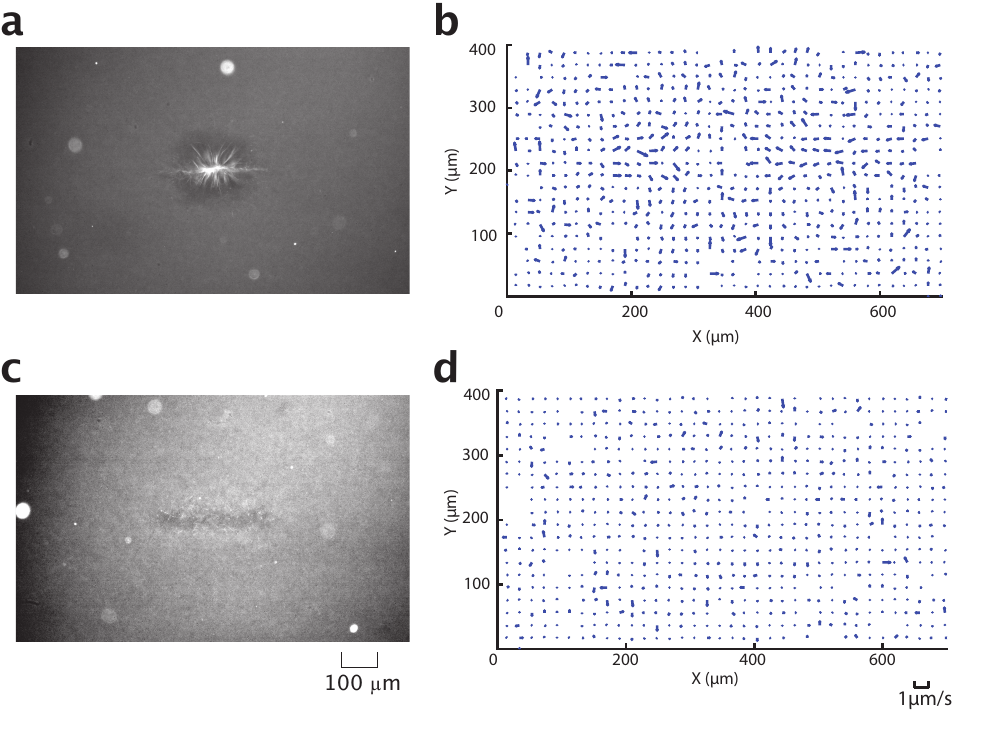}
    \caption{Minimum height experiment for a 350 x H $\upmu$m excitation pattern. \textbf{a}, Fluorescent microtubule channel for H $= 2$ $\upmu$m. \textbf{b}, Corresponding flow field to (\textbf{a}). \textbf{c}, Fluorescent microtubule channel for H $= 1$ $\upmu$m. \textbf{d}, Corresponding flow field to (\textbf{c}). }
    \label{fig:minheight}
\end{figure}

In addition, we find that the minimum height of an excitation bar that can generate flow is $\approx 2 \upmu$m Fig.~\ref{fig:minheight}. We observe that the network that forms is $\approx 300$ x $20$ $\upmu$m. Below this excitation limit we observe the formation of unstable microtubule bundles that do not persist long enough to form a more ordered structure. While the excitation bar extends 350 $\upmu$m, we speculate that below the minimum height, the density of active motors is too low to completely drive organization. This may be a result of the diffusivity and speed of the motor proteins.

\begin{figure}[H]
    \centering
    \includegraphics[width=.95\textwidth]{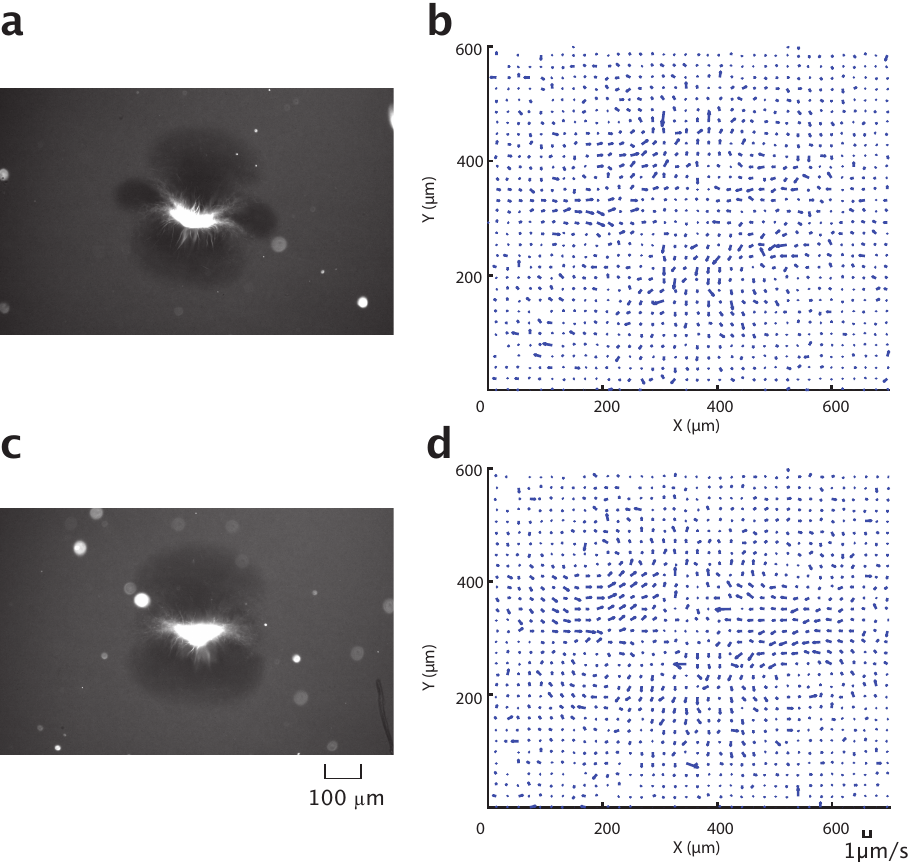}
    \caption{Minimum angle experiment for two 350 x 20 $\upmu$m excitation pattern. \textbf{a}, Fluorescent microtubule channel for an angle of $\frac{\pi}{8}$. \textbf{b}, Corresponding flow field to (\textbf{a}). \textbf{c}, Fluorescent microtubule channel for an angle of $\frac{\pi}{16}$. \textbf{d}, Corresponding flow field to (\textbf{c}).}
    \label{fig:minangle}
\end{figure}

We determine the angle resolution by taking two overlapping bars, as in the ``+'' shape shown in Fig.~4f, and rotating them relative to each other. When the bars are orthogonal to each other, there are four distinct inflows at the corners. We decrease the angle between the bars until the flow pattern appears to be that of a single bar (two inflows). The minimum angle between two bar patterns for which there remain 4 distinct inflows and outflows is between $\frac{\pi}{16} - \frac{\pi}{8}$ Fig.~\ref{fig:minangle}. The angle that sets this limit may in part be set by the average length of the filament bundles that form orthogonal to the major axis of each bar pattern are $\approx 20$ $\upmu$m in length. For a sufficiently shallow angle, these orthogonal bundles may interact with each other and cause the two microtubule networks to be pulled into each other, merging into a single linear structure.  The flow pattern and microtubule distribution of Fig.~\ref{fig:minangle}\textbf{c} and \textbf{d} closely resemble those produced by a single rectangular bar of light. 

\begin{figure}[H]
    \centering
    \includegraphics[width=.7\textwidth]{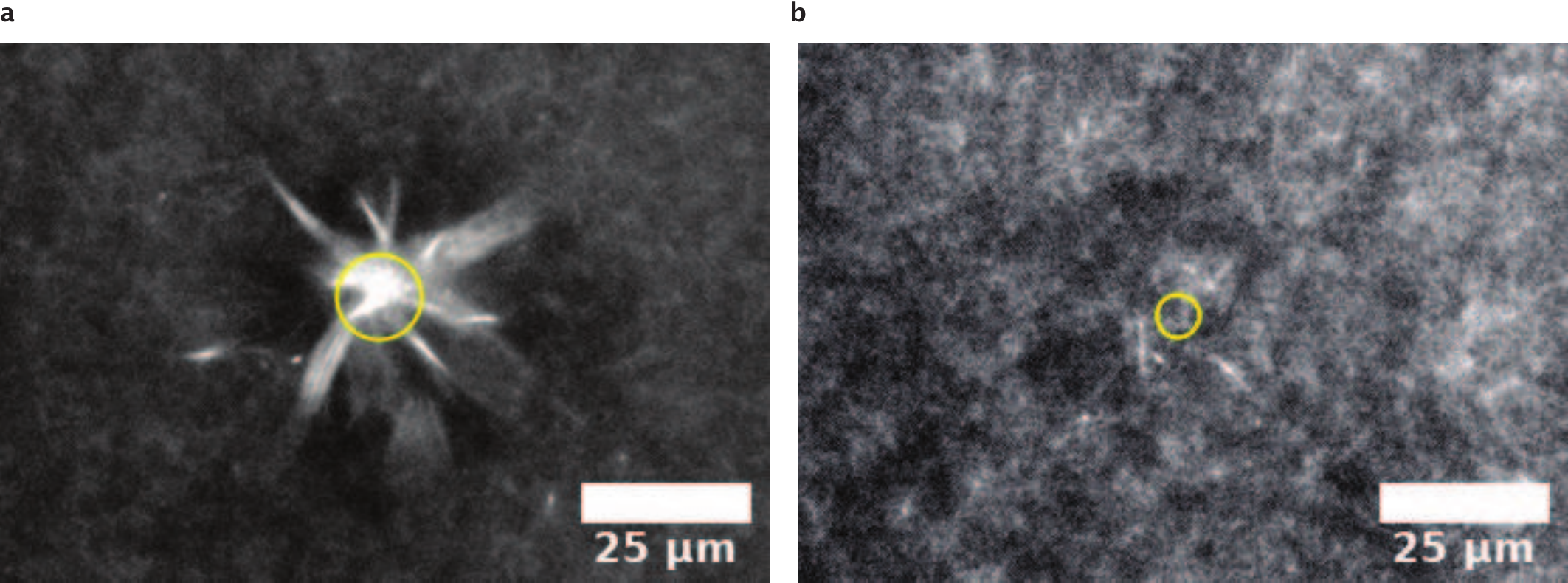}
    \caption{Minimum aster size experiment for disk patterns. \textbf{a}, Fluorescent microtubule channel for an excitation disk $12.5$ $\upmu$m excitation disk. \textbf{b}, Fluorescent microtubule channel for a $6.25$ $\upmu$m excitation disk. The yellow circle represents the perimeter of the excitation disk.}
    \label{fig:mindisk}
\end{figure}

 We find that the minimum disk diameter to form an aster is between $6.25-12.5$ $\upmu$m Fig.~\ref{fig:mindisk}. The arms of the smallest aster we are able to form appear to be $\approx 20$  $\upmu$m. We note that below this limit, small microtubule bundles form transiently and remain disordered. Due to the similarity of the minimum excitation length scale to the average microtubule length, we hypothesize that the smallest aster we can form may in part be determined by the microtubule length distribution.

\subsection{Fluid Flow Patterns from Particle Tracking}\label{dsec:flowtrack}
The fluid flow generated by the movement of microtubule filaments is measured using Particle Tracking Velocimetry (PTV) \cite{maas1993particle} of fiducial tracer particles. Inert 1 $\upmu$m diameter microspheres (SI~\ref{msec:tracerparticles}) are added to the reaction buffer and imaged with brightfield microscopy. The images are pre-processed using a Gaussian filter with a standard deviation of $1$ pixel, followed by thresholding to eliminate the background noise. After filtering, the centroid of each particle is measured and tracked. 

\begin{figure}[H]
    \centering
    \includegraphics[width=.9\textwidth]{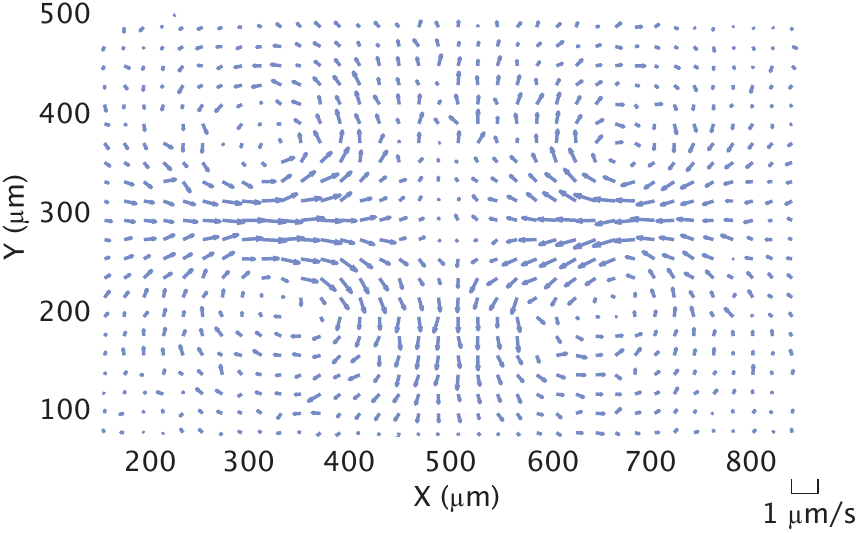}
    \caption{Flow velocity field generated with a 350 $\upmu$m activation bar measured with PTV of tracer particles. Vector data is used to calculate streamline plot in Fig.~4c.}
    \label{fig:flowvelocity}
\end{figure}

A nearest-neighbor algorithm \cite{schmidt1996imaging} is applied to find particle pairs within a square search window (30 pixels). Displacement vectors are then calculated by comparing the position of particle pairs in consecutive frames. The same process is repeated for the entire image sequence ($30$ min). The velocity field is generated by dividing the displacement vector field by the time interval between frames. The averaged velocity field shown in Fig.~\ref{fig:flowvelocity} is carried out by grouping and averaging all velocity vectors within a $30$ pixel $\times 30$ pixel window.

\subsection{2D Flow Field}\label{dsec:2Dflow}
\begin{figure}[H]
    \centering
    \includegraphics[width=0.9\textwidth]{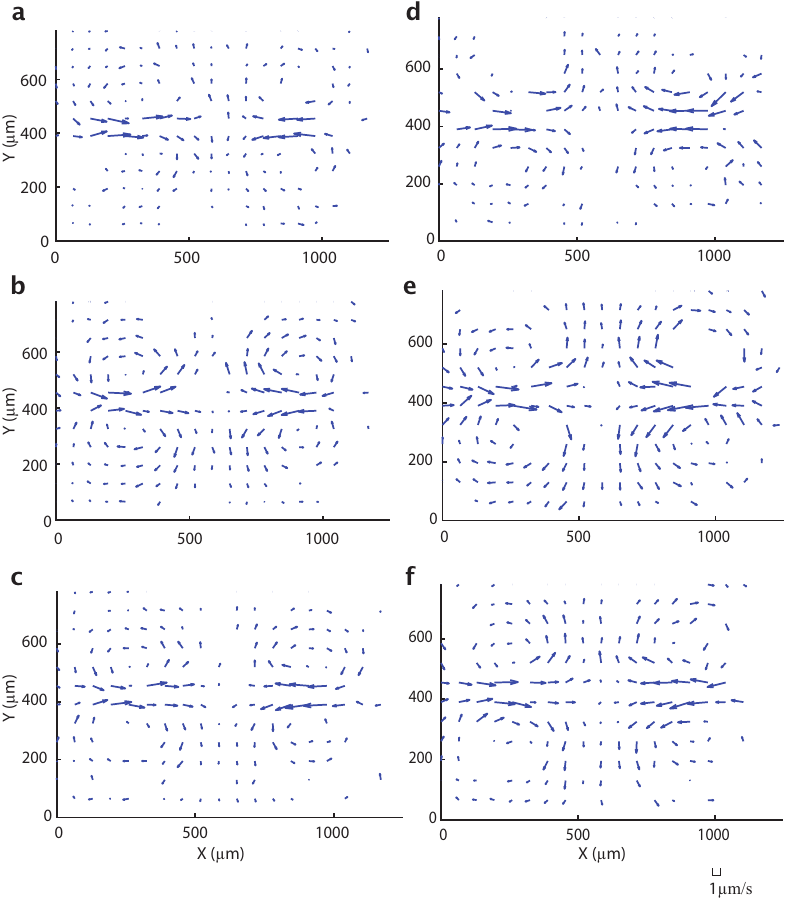}
    \caption{A flow field measured at three different z-positions separated by 20 $\upmu$m. The field is generated with a 700 $\upmu$m activation bar. \textbf{a}, Highest z-position, \textbf{b}, middle z-position, \textbf{c}, lowest z-position. \textbf{d}, \textbf{e}, \textbf{f}, are from another experiment following the same order.}
    \label{fig:zstack}
\end{figure}

We measure the flow field at different focal planes to determine its z-dependence. The flow fields are generated from PTV, as previously described (SI~\ref{dsec:flowtrack}). We image a z-stack of 3 planes separated by 20 $\upmu$m, where the sample typically extends $\approx 70  \upmu$m in the z-direction. Following the same particle tracking algorithm, we retrieve the flow fields (Fig.~\ref{fig:zstack}) averaged over a 20 min time window. We do not observe significant differences in the flow field's structure or speed at the various z-positions. Therefore, for all subsequent flow measurements we image a single focal plane.  Further, when we model the flow field (SI~\ref{dsec:stokeslets}), we assume it is a 2D pattern. 

\subsection{Time Stability of Flow Patterns}\label{dsec:flowpatterns}
\begin{figure}[H]
    \centering
    \includegraphics[width=.9\textwidth]{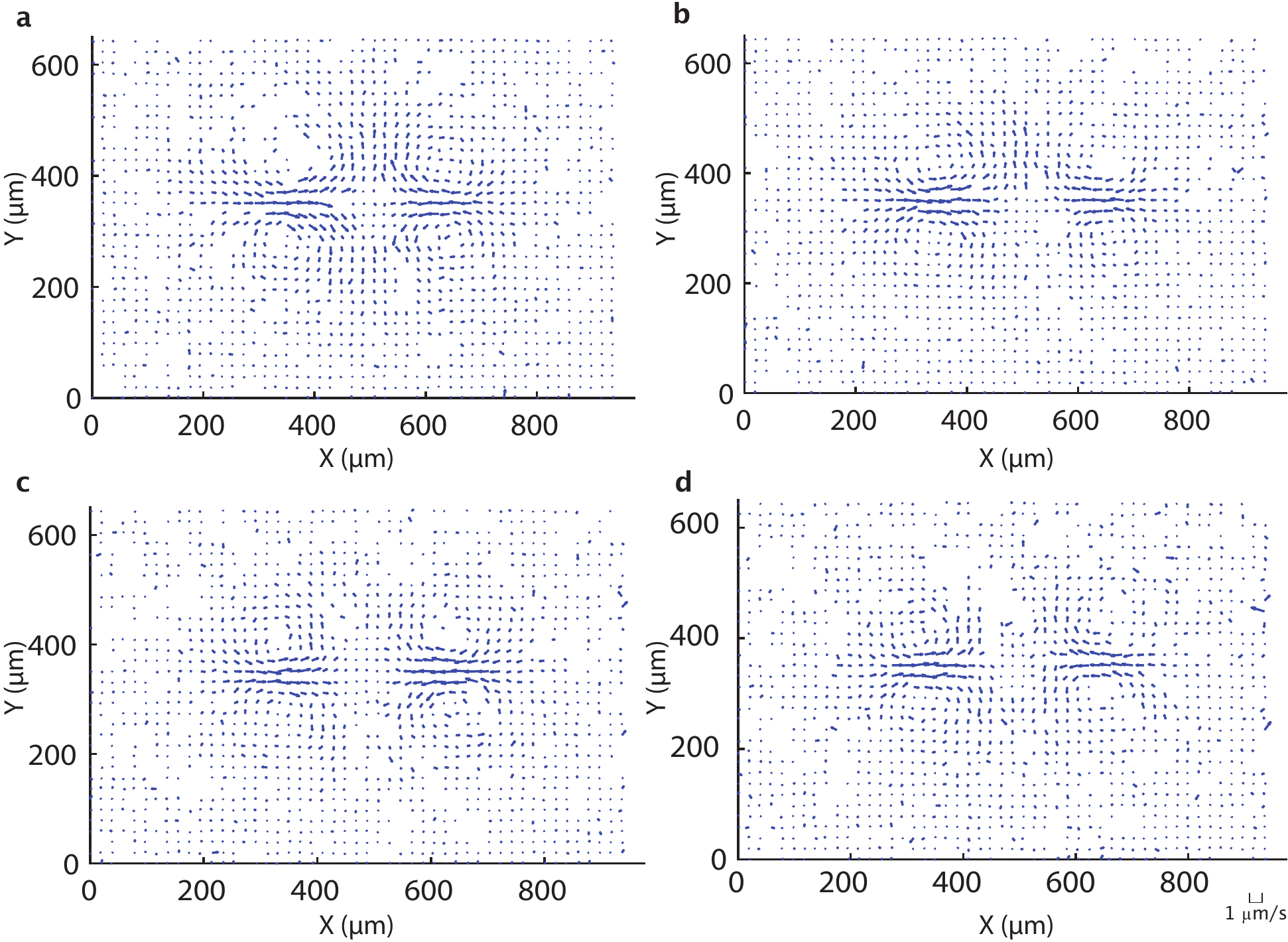}
    \caption{Velocity field averaged over $7.5$ minute intervals in a single experiment. Time windows are \textbf{a}, t = 0 - 7.5 min \textbf{b}, t = 7.5 - 15 min \textbf{c}, t = 15 - 22.5 min \textbf{d}, t = 22.5 - 30 min}
    \label{fig:flowfield}
\end{figure}

\begin{figure}[H]
    \centering
    \includegraphics[width=.5\textwidth]{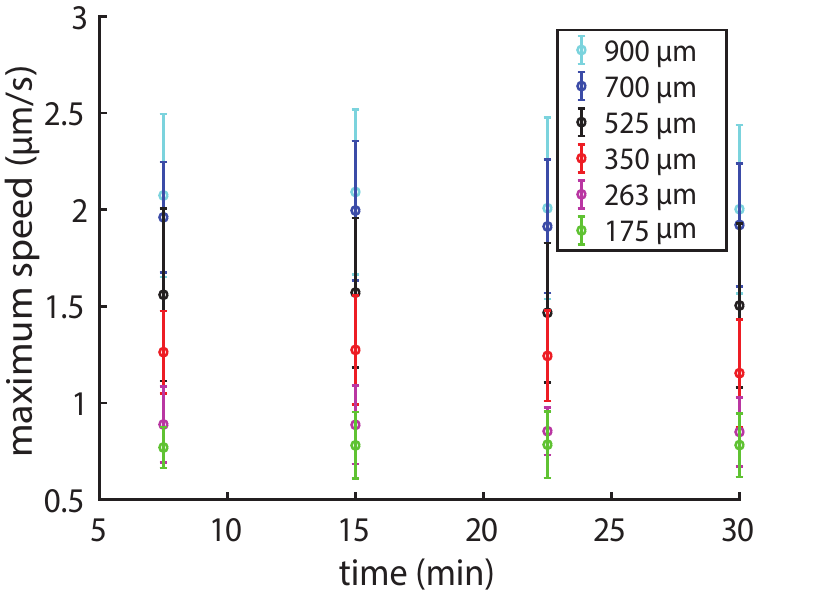}
    \caption{The average maximum speed for four different $7.5$ minute time windows. The data points represent the average of nine experiments. The error bars are the associated standard deviation.}
    \label{fig:speed}
\end{figure}

In order to understand how the flow field changes in time, we divide the 30 minute experiment into four 7.5 minute time windows and calculate the flow field for each window. The resulting velocity fields are shown in (Fig.~\ref{fig:flowfield}). We note that the structure of the flow field remains similar throughout the experiment. In addition, the maximum speed of the velocity field is constant over time (Fig.~\ref{fig:speed}), which further confirms that the fluid flow is stable over the experiment.

\subsection{Generation of Streamline Plots}\label{dsec:streamline}
Streamlines are the spatial path traced out by fiducial points moving with the fluid flow.  They can be numerically generated from a velocity vector field. To generate the streamlines shown in Fig.~4c, g we use the streamplot function found in the Matplotlib Python library. First, the streamplot function maps a user-defined grid onto the velocity vector field, which determines the density of the streamlines. Next, streamplot creates trajectories from a subset of velocity vectors by performing an interpolation from the current position $x(t)$ of the streamline to the next position $x(t+\mathrm{d}t)$ based on the velocity $v(x(t))$ by a 2nd-order Runge-Kutta algorithm.  To prevent streamlines from crossing, a mask is defined around each interpolated trajectory, which excludes other trajectories from entering into the mask.

\subsection{Correlation Length}\label{dsec:correlation}
\begin{figure}[H]
    \centering
    \includegraphics[width=.5\textwidth]{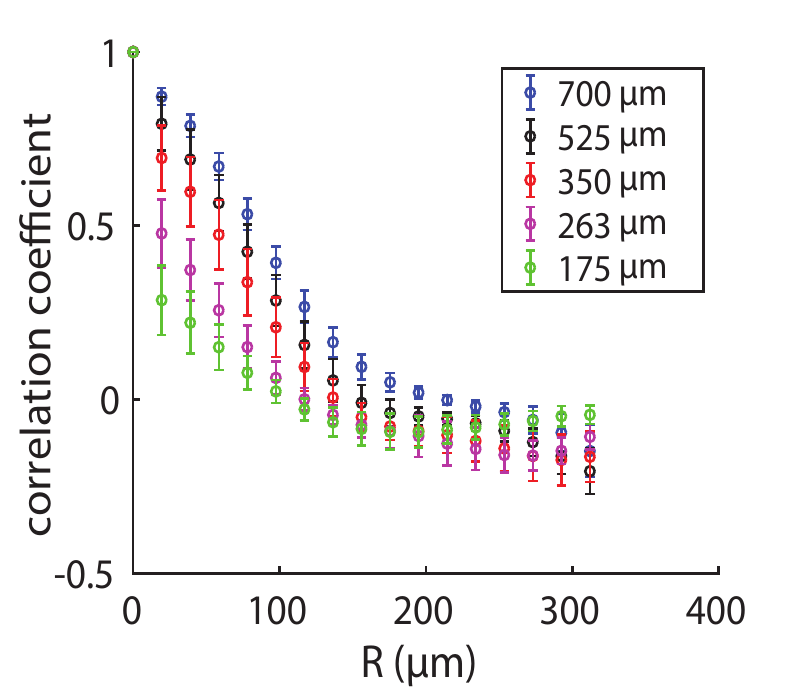}
    \caption{The correlation coefficient as a function of distance. Each marker shows the mean over nine individual experiments and error bars are the associated standard deviation.}
    \label{fig:cc}
\end{figure}

The flow patterns that we observe have vortices.  We can characterize the spatial extent of patterns like vortices by the velocity–velocity correlation coefficient $C(R)$  \cite{dunkel2013fluid, sanchez_spontaneous_2012}:
\begin{equation}
    C(R) = \frac{\left<V(R)\cdot V(0)\right>}{\left<|V(0)|^{2}\right>}
\end{equation}
where $V$ is the fluid velocity vector, $R$ is the distance between velocity vectors, $\left<\,\right>$ denotes assemble average and || is the magnitude of the vector. The correlation length $L_{c}$ is defined as the distance when $C(L_{c}) = 0$. This is the length scale where velocities vectors change to an orthogonal direction. By definition, $C(0) = 1$. The correlation coefficient as a function of $R$ is calculated to determine  $L_{c}$ for each bar length (Fig.~\ref{fig:cc}).

\subsection{Theoretical Model of the Fluid Flow Field}\label{dsec:stokeslets}
\begin{figure}[H]
    \centering
    \includegraphics[width=0.7\textwidth]{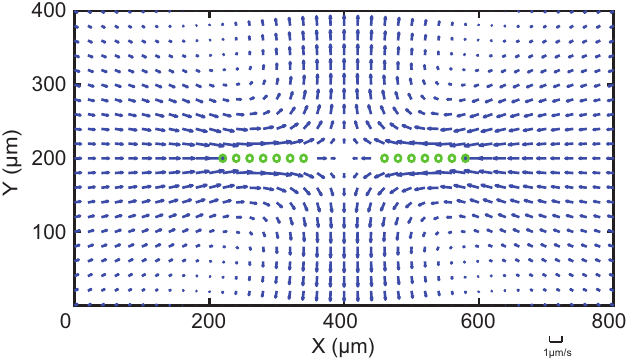}
    \caption{Flow field generated by 14 Stokeslets, indicated by green circles, to model the 350 $\upmu$m activation bar case. This theoretical model recovers the general pattern of inflows and outflows observed in the experiment (Fig.~4a), but not the vortices and asymmetries in flow magnitudes.}
    \label{fig:model}
\end{figure}

We use solutions of the Stokes equation, the governing equation for fluid flow at low-Reynolds-number \cite{happel2012low}, to model our induced flow fields. One of the simplest solutions of the equation is the Stokeslet, which describes the flow field induced by a point force \cite{chwang1975hydromechanics}. Here, we attribute the flow-generating point forces to contracting microtubule bundles.  Since the microtubules at the center of the activation bar appear to contract much more slowly than in other regions of the light pattern, we do not model Stokeslets in the central 120 $\upmu$m of the activation bar. We superimpose the solutions for two series of Stokeslets, one for each side of the bar. Each series of Stokeslets is composed of 7 point forces with identical magnitude (|$\bm{f}$| = 2 nN), separated by 20 $\upmu$m (Fig.~\ref{fig:model}) to model the 350 $\upmu$m activation bar case.

The velocity field $\bm{u}$($\bm{x}$) generated by a point force $\bm{f}$ located at $\bm{x'}$ in a 2D plane is given as
\begin{equation}
    \bm{u}(\bm{x}) = \frac{1}{4\pi \eta}\left(-\bm{f}\log(r) + \frac{(\bm{f} \cdot (\bm{x-x'}))(\bm{x-x'})}{r^{2}}\right)
\end{equation}
where $\eta$ is the fluid viscosity and $r$ is the absolute distance, defined as
\begin{equation}
    r = |\bm{x-x'}|.
\end{equation}
We estimate $\eta$ = $2 \times 10^{-3}$ $\text{Pa} \cdot \text{s}$ (Supplementary Information~\ref{dsec:viscosity}).

Comparing Fig.~\ref{fig:model} to Fig.~\ref{fig:flowvelocity}, for the rectangular bar experiment, we see our model recovers the general pattern of inflows and outflows in magnitude and direction. In both figures, the inflows along the X direction and the outflows along the Y direction are asymmetric in magnitude, with the inflows being greater than the outflows. However, in the experiments there can be additional asymmetries not captured by the model.  For example in Fig.~\ref{fig:flowvelocity}, outflows in the downward direction (Y-axis, Y < 300 $\upmu$m) appear greater in magnitude than the outflows in the upward direction (Y-axis, Y > 300 $\upmu$m).  This may be related to the microtubule buckling shown in Fig. 4b, which leads to asymmetry of the microtubule network density in the last panel of Fig. 4a.  Further, we note that we do not observe vortices for our model parameters.  It is possible that the presence of vortices may lead to additional effects not generated by the current model. 

There are various candidate mechanisms for vortex generation - boundary conditions, zones of depleted microtubules, and non-Newtonian fluid properties, to list a few.  Further investigation will be needed to determine which of these effects, if any, cause the observed vortices.  

\begin{figure}[H]
    \centering
    \includegraphics[width=1\textwidth]{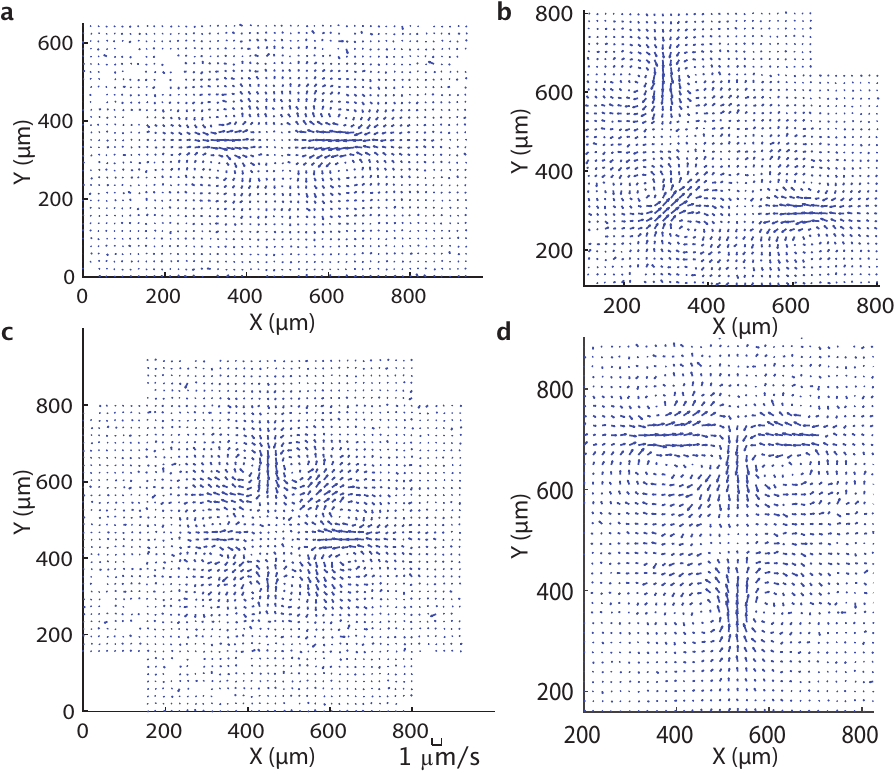}
    \caption{Demonstration of the linearity of the flow field. \textbf{a}, A time averaged flow field generated by a 350 $\upmu$m rectangular bar. Flow fields generated by the rotation and superposition of the flow field in (\textbf{a}) to retrieve flow fields for \textbf{b}, ``L'' \textbf{c}, ``+'', and \textbf{d}, ``T''-shaped light patterns.}
    \label{fig:superposition}
\end{figure}

Due to the linear nature of low-Reynolds-number flow  \cite{kim2013microhydrodynamics}, we expect that the velocity field generated by a complex light pattern can be retrieved by superposition of simple patterns. To confirm this, we superimpose flow fields from single bars to mimic the flow field generated by ``L'', ``+'' and ``T''-shaped light patterns (Fig.~\ref{fig:superposition}). For the ``+'' case, the superimposed fields closely resembles the experimentally observed field (Fig.~\ref{fig:superposition}c). The ``L'' and ``T''-shaped cases are roughly similar to the experimental results, but direction of the inflows do not match (Fig.~\ref{fig:superposition}b, d). 

\begin{figure}[H]
    \centering
    \includegraphics[width=1\textwidth]{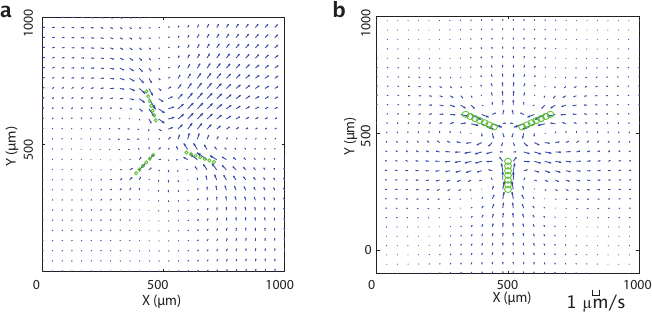}
    \caption{Theoretical simulation of fluid flows under complex light patterns using Stokeslets. The Stokeslets are positioned following the shape of the microtuble network observed in Fig.~4f. Green circles denote the Stokeslets. \textbf{a}, Flow field for ``L''-shaped light pattern. \textbf{b}, Flow field for ``T''-shaped light pattern.}
    \label{fig:simulation_complex}
\end{figure}

To model the ``L'' and ``T'' flow fields more accurately, we generate the flow field for a series of Stokeslets following the geometry of the microtubule structure, rather than the light pattern itself. Using this method, the modeled flow fields are a good approximation of the observed flow fields. The inflows and outflows match the experimentally observed positions and orientations (Fig.~\ref{fig:simulation_complex}). This result implies that the observed flow patterns are set by the microtubule structure rather than the light pattern.

\subsection{Calculating Fluid Viscosity}\label{dsec:viscosity}
To find the viscosity of the background buffer, we used a similar approach to finding the flow fields.  We used PTV of fiducial tracer particles (Supplemental Information~\ref{dsec:flowtrack}) in inactivated regions of the sample of the 175 $\upmu$m activation bar experiment. Assuming the buffer is Newtonian \cite{panton2006incompressible}, the inert tracer particles diffuse freely due to thermal fluctuations. From the tracking results, we measure the mean-squared displacement MSD$(t)$ of the particles:
 \begin{equation}
     \text{MSD}(t) = \left<(x(t) - x(0))^2+(y(t) - y(0))^2\right>,
 \end{equation}
where $x(t)$ and $y(t)$ are the position of a given particle at time $t$ and $\left<\,\right>$ denotes ensemble average. For this calculation, each frame is $t =$ 4 s apart. The MSD$(t)$ of a freely diffused particle in 2D follows the Stokes-Einstein equation
\begin{equation}
    \text{MSD}(t) = 4Dt = \frac{2k_BT}{3\pi\eta r}t,
\end{equation}
where $r = 0.5 \, \upmu$m is the radius of the particle. Then, the viscosity of the buffer solution is estimated as 
\begin{equation}
    \eta = \frac{8k_BT}{3\pi r \text{MSD}(t)}.
\end{equation}
The same process is repeated through nine individual experiments and the average estimated viscosity $\eta$ is $2 \times 10^{-3}$ $\text{Pa} \cdot \text{s}$.

\subsection{Comparison to Optically Controlled Bacteria}\label{dsec:swimsys}

The polarity of the motors and microtubules makes them distinct from systems based on optically controlled bacteria \cite{frangipane_dynamic_2018,arlt_painting_2018}. In our work, the localization of motor linkages causes microtubules to collectively reorganize into contracting networks. Due to the organization of microtubules and resulting dipolar stresses on the surrounding medium, we are able to create coherent flows. In contrast, localization of the activity of bacterial swimmers results in a change in the bacterial density, but lacks structural order and therefore does not generate coherent flows. However, bacterial densities can form arbitrary patterns that directly correspond to the optical projections analogous to photolithography. The resolution of the patterns we can create (Supplementary Information~\ref{dsec:minsize}) is generally lower than the reported $\approx$ 2 $\upmu$m resolution achievable with bacterial swimmers. Light in our system does not directly pattern microtubules but rather defines an effective reaction volume where certain reorganizing motifs can occur. 
\printbibliography

\end{document}